\documentclass[12pt]{article}%
\IfFileExists{sariel_computer.sty}{\def\sarielComp{1}}{}
\ifx\sarielComp\undefined%
\newcommand{\SarielComp}[1]{}
\newcommand{\NotSarielComp}[1]{#1}%
\else
\newcommand{\SarielComp}[1]{#1}%
\newcommand{\NotSarielComp}[1]{}%
\fi
\newcommand{\IfPrinterVer}[2]{#2}%

\usepackage{caption}%
\usepackage{comment}%
\usepackage[cm]{fullpage}%
\usepackage{amsmath}%
\usepackage{amssymb}%
\usepackage{xcolor}%

\SarielComp{\usepackage{sariel_colors}}%

\usepackage[amsmath,thmmarks]{ntheorem}%
\theoremseparator{.}%

\usepackage{titlesec}%
\titlelabel{\thetitle. }%
\usepackage{xcolor}%
\usepackage{mleftright}%
\usepackage{xspace}%
\usepackage{hyperref}%
\usepackage{graphicx}%

\newcommand{\hrefb}[3][black]{\href{#2}{\color{#1}{#3}}}%

\IfPrinterVer{%
   \usepackage{hyperref}%
}{%
   \usepackage{hyperref}%
   \hypersetup{%
      breaklinks,%
      ocgcolorlinks, colorlinks=true,%
      urlcolor=[rgb]{0.25,0.0,0.0},%
      linkcolor=[rgb]{0.5,0.0,0.0},%
      citecolor=[rgb]{0,0.2,0.445},%
      filecolor=[rgb]{0,0,0.4},
      anchorcolor=[rgb]={0.0,0.1,0.2}%
   }
}

\theoremseparator{.}%

\theoremstyle{plain}%
\newtheorem{theorem}{Theorem}[section]

\newtheorem{lemma}[theorem]{Lemma}

\theoremstyle{plain}%
\theoremheaderfont{\sf} \theorembodyfont{\upshape}%
\newtheorem*{remark:unnumbered}[theorem]{Remark}%
\newcommand{\myqedsymbol}{\rule{2mm}{2mm}}

\theoremheaderfont{\em}%
\theorembodyfont{\upshape}%
\theoremstyle{nonumberplain}%
\theoremseparator{}%
\theoremsymbol{\myqedsymbol}%

\newcommand{\HLink}[2]{\hyperref[#2]{#1~\ref*{#2}}}
\newcommand{\HLinkSuffix}[3]{\hyperref[#2]{#1\ref*{#2}{#3}}}

\newcommand{\apndlab}[1]{\label{apnd:#1}}
\newcommand{\apndref}[1]{\HLink{Appendix}{apnd:#1}}

\newcommand{\figlab}[1]{\label{fig:#1}}
\newcommand{\figref}[1]{\HLink{Figure}{fig:#1}}

\newcommand{\thmlab}[1]{{\label{theo:#1}}}
\newcommand{\thmref}[1]{\HLink{Theorem}{theo:#1}}

\newcommand{\seclab}[1]{\label{sec:#1}}
\newcommand{\secref}[1]{\HLink{Section}{sec:#1}}

\newcommand{\lemlab}[1]{\label{lemma:#1}}
\newcommand{\lemref}[1]{\HLink{Lemma}{lemma:#1}}%

\providecommand{\eqlab}[1]{}%
\renewcommand{\eqlab}[1]{\label{equation:#1}}
\newcommand{\Eqref}[1]{\HLinkSuffix{Eq.~(}{equation:#1}{)}}

\newcommand{\remove}[1]{}%

\renewcommand{\th}{th\xspace}

\renewcommand{\Re}{\mathbb{R}}%
\newcommand{\reals}{\Re}%

\providecommand{\emphi}[1]{\emph{#1}}%

\def\eps{{\varepsilon}}  \def\A{{\cal A}}

\newcommand{\etal}{\textit{et~al.}\xspace}

\numberwithin{figure}{section}%
\numberwithin{table}{section}%
\numberwithin{equation}{section}%

\newcommand{\LUp}{\Lambda^\uparrow}%
\newcommand{\LDown}{\Lambda^\downarrow}%
\newcommand{\Arr}{\mathcal{A}}%
\newcommand{\Saarbrucken}{Saarbr\"ucken\xspace}

\title{The Maximum-Level Vertex in an Arrangement of Lines}

\author{%
   Dan Halperin\thanks{%
      School of Computer Science, Tel Aviv University, Tel~Aviv 69978,
      Israel; {\tt danha@tau.ac.il}.}  \and %
   Sariel Har-Peled\thanks{%
      Department of Computer Science; University of Illinois; 201
      N. Goodwin Avenue; Urbana, IL, 61801, USA; {\tt
         sariel@illinois.edu}.}  \and %
   Kurt Mehlhorn\thanks{%
      Max Planck Institute for Informatics, Saarland Informatics
      Campus, 66123 \Saarbrucken, Germany; {\tt
         mehlhorn@mpi-inf.mpg.de}.}  \and Eunjin Oh\thanks{%
      Max Planck Institute for Informatics, Saarland Informatics
      Campus, 66123 \Saarbrucken, Germany; Present address: Department
      of Computer Science and Engineering, POSTECH, Pohang 37673,
      Korea; {\tt eunjin.oh@postech.ac.kr}.}  \and Micha
   Sharir\thanks{%
      School of Computer Science, Tel Aviv University, Tel~Aviv 69978,
      Israel; {\tt michas@tau.ac.il}.}
}%

\begin{document}

\maketitle

\vspace{-3cm}%
\centerline{\includegraphics[width=0.95\linewidth]%
   {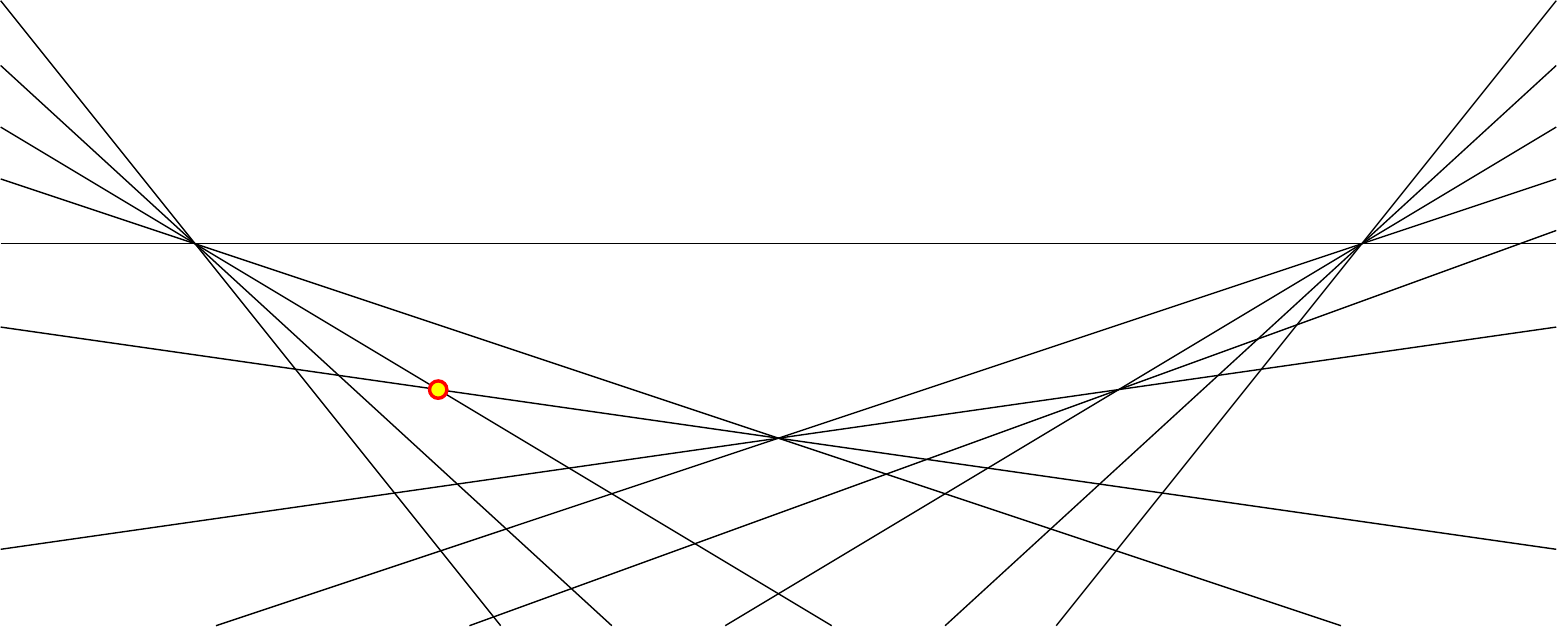}}

\begin{abstract}
    Let $L$ be a set of $n$ lines in the plane, not necessarily in
    general position. We present an efficient algorithm for finding
    all the vertices of the arrangement $\A(L)$ of maximum level,
    where the level of a vertex $v$ is the number of lines of $L$ that
    pass strictly below $v$. The problem, posed in Exercise~8.13 in de
    Berg \etal~\cite{bcko-08}, appears to be much harder than it
    seems,
    as this vertex might not be on the upper envelope of the lines.
    
    We first assume that all the lines of $L$ are distinct, and
    distinguish between two cases, depending on whether or not the
    upper envelope of $L$ contains a bounded edge. In the former case,
    we show that the number of lines of $L$ that pass \emph{above} any
    maximum level vertex $v_0$ is only $O(\log n)$. In the latter
    case, we establish a similar property that holds after we remove
    some of the lines that are incident to the single vertex of the
    upper envelope.
    We present algorithms that run, in both cases, in optimal
    $O(n\log n)$ time.
    
    We then consider the case where the lines of $L$ are not
    necessarily distinct.  This setup is more challenging, and the
    best we have is an algorithm that computes all the maximum-level
    vertices in time $O(n^{4/3}\log^{3}n)$.

    Finally, we consider a related combinatorial question for
    degenerate arrangements, where many lines may intersect in a
    single point, but all the lines are distinct: We bound the
    complexity of the \emph{weighted $k$-level} in such an
    arrangement, where the weight of a vertex is the number of lines
    that pass through the vertex.  We show that the bound in this case
    is $O(n^{4/3})$, which matches the corresponding bound for
    non-degenerate arrangements, and we use this bound in the analysis
    of one of our algorithms.
\end{abstract}

\section{Introduction}
\seclab{intro}

Let $L$ be a set of $n$ lines in the plane, not necessarily in general
position (that is, there may be points incident to more than two lines
of $L$, and pairs of lines of $L$ might be parallel or even
coincide). The largest part of the paper is devoted to the case where
the lines of $L$ are pairwise distinct; the more difficult case where
lines of $L$ might coincide will be handled later on. We wish to find
a vertex, or rather all the vertices, of the arrangement $\A(L)$ at
maximum level, where the level $\lambda(v)$ of a vertex $v$ is the
number of lines of $L$ that pass strictly below $v$.

The question that we address here appears as an exercise in the
computational geometry textbook by de Berg \etal
\cite[Exercise~8.13]{bcko-08}. It can be solved in quadratic time by
constructing the full arrangement, and then by tracing the vertices
along each line from left to right, keeping track of the level of each
vertex as we go. The challenge is of course to solve it faster.

If we assume general position (so no three lines pass through a common
point), then every vertex on the upper envelope of $L$ is at level
$n-2$, which is the maximum possible level (and only the vertices of
the envelope have this level). Finding one such vertex in linear time
is straightforward,\footnote{%
   For example, compute the top line $\ell$ intersecting the $y$-axis,
   and then compute the at most two consecutive vertices of the
   arrangement along $\ell$ adjacent to this intersection.}  and
finding all of them takes $O(n\log n)$ time. Henceforth we focus on
the interesting, and harder, case where the lines are not in general
position. For this setting we are not aware of any previous
subquadratic-time algorithm to compute a maximum-level vertex. As the
requirement of Exercise 8.13\ in \cite{bcko-08} was to solve the
problem in $O(n\log n)$ time, it seems that the difficulty of the
problem was overlooked there.

The main obstacle is that, in degenerate situations, the desired
vertex does not have to lie on the upper envelope of $L$, as shown in
the example depicted in \figref{not:on:env}.

\begin{figure}[h tb]
    \begin{tabular}[t]{cc}
      \begin{minipage}{0.35\linewidth}
          \centerline{{\includegraphics{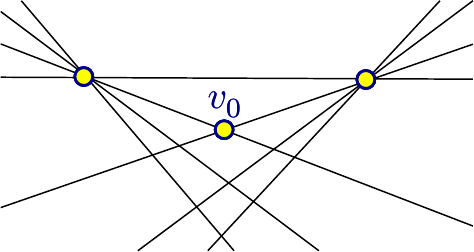}}}%
      \end{minipage}
      & 
        \begin{minipage}{0.55\linewidth}
            \centerline{\includegraphics{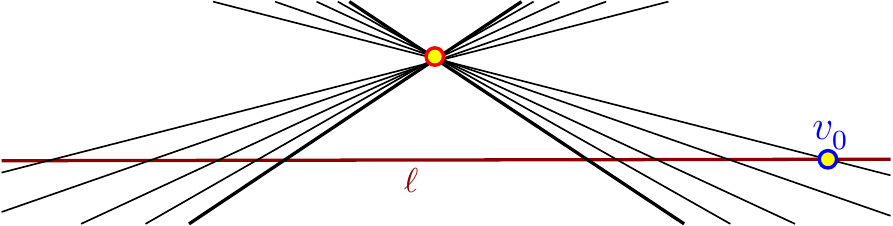}}
        \end{minipage}
      \\[0.2cm]
      \begin{minipage}{0.35\linewidth}
          \caption{\sf A set $L$ of lines for which the vertex of
             $\A(L)$ of maximum level, which is $v_0$, does not lie on
             the upper envelope.}%
          \figlab{not:on:env}
      \end{minipage}
      &
        \quad\begin{minipage}{0.55\linewidth}
            \caption{\sf A more ``substantial'' construction, in which
               the horizontal line $\ell$ contains all but one of the
               vertices of $\A(L)$, all at upper level
               $\Theta(n)$. The maximum-level vertex is $v_0$ (as well
               as its symmetric counterpart on the other side of
               $\ell$).}
            \figlab{very:no:env}
        \end{minipage}
    \end{tabular}
\end{figure}

In fact, the situation can be much worse---the vertex at maximum level
can be far away from the upper envelope. An illustration of such a
case is given in \figref{very:no:env}.

We do not solve exercise 8.13 completely. We give an $O(n \log n)$
algorithm only for the case of distinct lines. For the case where the
lines in $L$ are not necessarily distinct we only give an
$O(n^{4/3}\log^3 n)$ algorithm.  In either case, we may assume that
$L$ does not contain any vertical line: any such line is not counted
in the level of any point, and the only role of such lines is to
create new vertices of the arrangement. For any vertical line $\ell$,
the only relevant vertex is the highest intersection point of $\ell$
with other lines of $L$. It is straightforward\footnote{%
   The divide-and-conquer algorithm for computing the upper envelope
   (split the set of lines into two parts of equal size, compute the
   upper envelope of each, and merge by a scan along both envelopes)
   is readily extended to also compute the degrees of the vertices on
   the upper envelope.}  to find, in $O(n\log n)$ overall time, these
highest intersection points and their levels, for all vertical lines.
Therefore, in what follows, we can indeed assume that $L$ has no
vertical lines.

Consider in what follows the case where all the lines of $L$ are
distinct; as already noted, the case of coinciding lines is subtler and is discussed in
detail in~\secref{coincide}.

Similar to the case of vertices, a point $p$ in the plane is said to be at level $k$, if there are
exactly $k$ lines in $L$ passing strictly below $p$. The level of a
(relatively open) edge $e$ (resp., face $f$) of $\A(L)$ is the level
of any point of $e$ (resp., $f$).  The $k$-level of $\A(L)$ is the
closure of the union of the edges of $\A(L)$ that are at level
$k$. The \emph{at-most-$k$-level} of $\A(L)$, or $(\leq k)$-level, is
the closure of the union of the edges of $\A(L)$ at levels $j$,
$0\leq j\leq k$.  We denote the $k$-level as $\LDown_k$, and the
at-most-$k$-level as $\LDown_{\leq k}$.

In complete analogy, we define the \emph{upper level} of a vertex $v$
in $\A(L)$ (or of any point $v\in\reals^2$) to be the number of lines
of $L$ that pass strictly above $v$.  The $k$-upper level and the
$({\leq}k)$-upper level of $\A(L)$ are defined analogously to the
standard level, and are denoted as $\LUp_k$ and $\LUp_{\le k}$,
respectively.

We consider two complementary cases:

\medskip
\noindent{\bf Case (i):}
The upper envelope of $L$ contains a bounded edge, and thus has at
least two vertices; see \figref{not:on:env}.

\medskip
\noindent{\bf Case (ii):}
The upper envelope of $L$ does not contain a bounded edge, and thus
consists of a single vertex and two rays; see \figref{very:no:env}.

\medskip The main combinatorial results that provide the basis for our
algorithms are summarized in the following two theorems.

\begin{theorem}
    \thmlab{case1}%
    Let $L$ be a set of $n$ distinct lines in the plane that satisfies
    the assumption of Case (i). Then the upper level of any
    maximum-level vertex of $\A(L)$ is at most $2\log n$.
\end{theorem}

For Case (ii) we can achieve a similar property with some additional
preparation.  Specifically, let $v$ be the single vertex of the upper
envelope of $L$, let $L_v$ denote the set of the lines of $L$ that are
incident to $v$, and set $K := L\setminus L_v$. Assume that $K$ is
nonempty; if $K = \emptyset$ then $v$ is the only vertex of $\A(L)$,
which is clearly of maximum level (which is $0$). For each line
$\ell\in L_v$, let $\ell^-$ (resp., $\ell^+$) denote the portion (ray)
of $\ell$ to the left (resp., right) of $v$. Set
$L_v^- = \{ \ell^- \mid \ell\in L_v\}$ and
$L_v^+ = \{ \ell^+ \mid \ell\in L_v\}$.  Sort the rays of $L_v^-$
downwards, i.e., in increasing order of their slopes, and sort the
rays of $L_v^+$ also downwards, now in decreasing order of their
slopes.  Let $D^-$ (resp., $D^+$) denote the size of the largest
prefix of the rays of $L_v^-$ (resp., $L_v^+$) that do not intersect
any line of $K$ (and thus any other line of $L$), and put
$D := \min\{D^-, D^+\}$.  See \figref{k:l}.

\begin{figure}[h]
    \phantom{}%
    \hfill%
    \includegraphics[page=1]{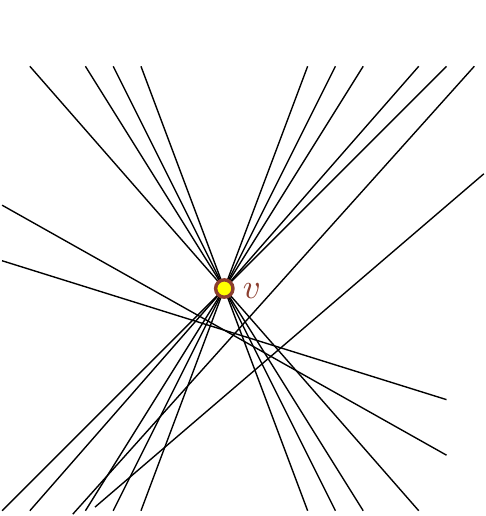}%
    \hfill%
    \includegraphics[page=2]{figs/k_l_decomp}%
    \hfill%
    \includegraphics[page=3]{figs/k_l_decomp} \hfill%
    \phantom{}%
    
    \caption{\sf The case of a single vertex on the upper envelope.
       The same arrangement is depicted three times, with
       different notations. }%
 \figlab{k:l}%
\end{figure}

Since $K\ne\emptyset$, it easily follows that no line $\ell$ of $L_v$
can contribute rays to both prefixes of $L_v^-$ and $L_v^+$ defined
above (unless all lines of $K$ are parallel to $\ell$, an easily
handled situation that we ignore here).

Put $h := \max\{0, D-2\log n\}$ and $D_0 := D-h = \min\{D,2\log n\}$.
Remove from $L$ the lines that contribute the $h$ topmost rays to
$L_v^-$ and the lines that contribute the $h$ topmost rays to $L_v^+$;
by what has just been said, no line is removed twice, and we are thus
left with a subset $L_0$ of $L$ of size $n-2h$.

\begin{theorem}
    \thmlab{case2}%
    Let $L$ be a set of $n$ distinct lines in the plane that satisfies
    the assumption of Case (ii). Let $v$, $L_v$, $K$, $D$, $h$, $D_0$
    and $L_0$ be as defined above.  Then all the maximum-level
    vertices of $\A(L)$ are vertices of $\A(L_0)$, and the upper level
    in $\A(L_0)$ of any maximum-level vertex of $\A(L)$ is at most
    $4 \log n$.
\end{theorem}

We will exploit these theorems in designing efficient algorithms, that
run in optimal $O(n\log n)$ time, for computing all the maximum-level
vertices, in both cases.  We note that this running time is indeed
optimal: Even the task of computing the upper envelope of $L$ is at
least as hard as the task of sorting the lines by slope.

A central ingredient of our algorithms is computing the
at-most-$k$-upper level of an arrangement,
where %
$k = O(\log n)$.  The complexity (number of edges and vertices) of the
$({\leq} k)$-level in an arrangement of $n$ lines is
$\Theta(nk)$~\cite{ag-86, cs-89}. Typically, this is shown for
arrangements of lines in general position, by an easy application of
the Clarkson-Shor random sampling theory~\cite{cs-89}, but it also
holds in degenerate situations, as can easily be verified.
The $({\leq}k)$-level can be computed, for arrangements in general
position, in optimal time $O(n\log n+nk)$ by an algorithm of Everett
\etal \cite{erk-96}. We sketch (our interpretation of) the algorithm
in \apndref{at:most:k}.
As $k= O(\log n)$ in both cases, the algorithm runs in (optimal)
$O(n\log n)$ time.
It is not clear, though, whether this (fairly involved) algorithm also
works for degenerate arrangements. %
   
To finesse this issue, we run the algorithm of \cite{erk-96} on
perturbed copies of the lines of $L$, using a simplified variant of
symbolic perturbation, and then extract from its output the actual
at-most-$k$-level in the original degenerate arrangement. In a fully
symmetric manner, this construction also applies to the
at-most-$k$-upper levels of $\A(L)$.

We remark that levels can be defined for arrangements of objects other
than lines and in higher dimensions. Levels in arrangements of
hyperplanes are closely related (by duality) to so-called \emph{$k$-sets} in
configurations of points. Both structures have been extensively
studied; see the recent survey on arrangements~\cite{hs-18} for a
review of bounds and algorithms.  In what follows, though, we only
concern ourselves with planar arrangements of lines.

The paper is organized as follows. In \secref{cases} we give the
proofs of \thmref{case1} and \thmref{case2}, and then present, in
\secref{algorithms}, our efficient (optimal) algorithms for both
cases. The case where $L$ can contain coinciding lines is discussed in
\secref{coincide}, where we present an algorithm that has a weaker
$O(n^{4/3}\log^{3}n)$
upper bound on its complexity.
We conclude in \secref{weighted} with a bound on the maximum
complexity of the \emph{weighted} $k$-level in arrangements of lines,
still catering to the case where many lines may intersect in a single
point, but the lines are all distinct.  Here the weight of a vertex is
the number of lines that pass through it, and the complexity of the
weighted level is the sum of the weights of its vertices.  On top of
being a result of independent interest, we exploit it in the analysis
of our algorithm for the case of coinciding lines.
In the Appendix we give a brief review of the optimal-time algorithm
by Everett \etal~\cite{erk-96} for computing the $({\leq}k)$-level for
arrangements of lines in general position, describing it from a
different (and, to us, simpler) perspective than the original paper.
\section{The upper level of maximum-level vertices}
\seclab{cases}%

The proofs of both \thmref{case1} and \thmref{case2} rely on the
following structural property, which we regard as interesting in its
own right.

Consider the $k$-upper level $\LUp_k$, which, as we recall, is the
$x$-monotone polygonal curve which is the closure of the union of the
edges of the arrangement with exactly $k$ lines above each of them.
Since the lines of $L$ are distinct, these levels do not share any
edge, but they can share vertices. The \emph{degree} of a vertex is
the number of lines in $L$ incident to the vertex. A vertex of degree
$d$ appears in $d$ consecutive levels. Note that the level does not
necessarily turn at every vertex $v$ that it reaches: it could pass
through $v$ staying on the same line (this happens when the degree of
$v$ is odd and the level reaches $v$ along the median incident
line). See \figref{does-not-turn} for an illustration.

\begin{figure}[h]
    \centerline{\includegraphics{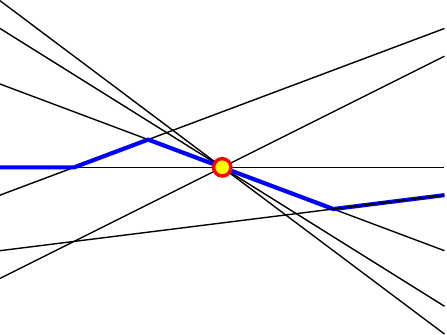}}%
    
    \caption{\sf The highlighted level does not turn at the marked vertex.}
    \figlab{does-not-turn}
\end{figure}

Let $k_0$ be the smallest index such that there exists some vertex $v$
that lies strictly above $\LUp_{k_0}$ (so $v$ is a vertex of
$\LUp_{k_0-1}$, but not necessarily of all the preceding upper
levels). The vertices lying strictly above $\LUp_{k_0}$ are called \emph{detached}. See \figref{k:0}.

\begin{figure}[h]
    \hfill%
    \includegraphics[page=2]{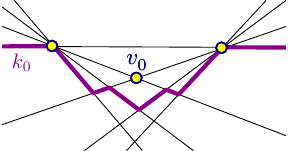}%
    \hfill%
    \includegraphics[page=2]{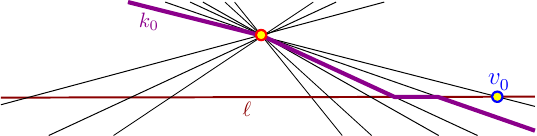}%
    \hfill%
    \phantom{}%
    \caption{\sf \figref{not:on:env} and a variant of \figref{very:no:env} with the
       $k_0$-upper level highlighted.}
    \figlab{k:0}
\end{figure}

\begin{lemma}%
    \lemlab{findv0}%
   A vertex has maximum level if and only if it lies above $\LUp_{k_0}$. The maximum level is $n - k_0$.
\end{lemma}
\noindent{\bf Proof.}
Let $v$ be any detached vertex. We claim that the level
$\lambda(v)$ of $v$ is exactly $n-k_0$. This is because there are
exactly $k_0$ lines that pass through or above $v$, which follows
since (i) this is the number of lines that cross the vertical line
through $v$ above $\LUp_{k_0}$, and (ii) none of these lines passes
between $v$ and $\LUp_{k_0}$, by definition.

Except for potential other vertices that lie, like $v$, strictly above
$\LUp_{k_0}$, and whose level is thus also $n-{k_0}$, any other vertex
$w$ lies on or below $\LUp_{k_0}$.  Suppose that $w$ lies on
$\LUp_{k_0}$. Move from $w$ slightly to its left, say, along an
adjacent edge of $\LUp_{k_0}$. The new point $w'$ has exactly ${k_0}$
lines above it and exactly one line through it, so its level satisfies
$\lambda(w') = n-{k_0}-1$. This implies that $\lambda(w)$ is at most
$n-{k_0}-1$, as we clearly must have $\lambda(w) \le \lambda(w')$; see
\figref{wandwp}.  The case where $w$ lies on an upper level of a
larger index is handled similarly, and in fact its level can only get
smaller. This completes the proof.  $\Box$

\medskip

\begin{figure}[h]
    \centerline{\includegraphics{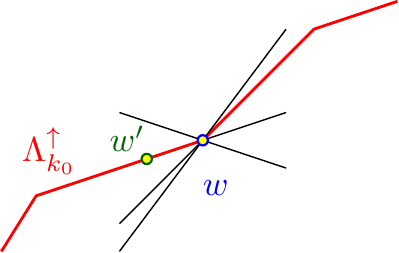}}%
    \caption{\sf The level of any vertex $w$ of $\LUp_{k_0}$ is at
       most $n-k_0-1$.}
    \figlab{wandwp}
\end{figure}

\medskip To exploit this result, we need the following property.
\begin{lemma} \lemlab{double} Assume that, for some $k\ge 0$, $\LUp_k$
    has at least two vertices, and that all the vertices of $\LUp_k$
    also belong to $\LUp_{k+1}$. Then, denoting by $V_j$ the number of
    vertices of $\LUp_j$, for any $j$, we have $V_{k+1} \ge 2V_k-1$.
\end{lemma}
\noindent{\bf Proof.}
The claim follows trivially by observing that if $a$ and $b$ are two
consecutive vertices of $\LUp_k$, and thus also of $\LUp_{k+1}$, then
$\LUp_{k+1}$ must contain at least one additional vertex\footnote{%
   Note that the assumption that the lines of $L$ are all distinct is
   crucial for this argument to apply.}  between $a$ and $b$. See
\figref{leveldoubles}.  Indeed, $\LUp_{k+1}$ leaves $a$ (to the right)
on a different edge than $ab$. Similarly, $\LUp_{k+1}$ enters $b$
(from the left) on a different edge than $ab$. These two edges must be
distinct, which implies that there must be at least one vertex in
between them on $\LUp_{k+1}$.  $\Box$

\begin{figure}[h]
    \centerline{\includegraphics{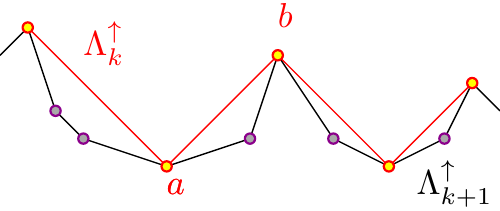}}
    \caption{\sf Proof of \lemref{double}. Any pair of consecutive
       upper levels $\LUp_k, \LUp_{k+1}$, such that all the vertices of
       $\LUp_k$ are also vertices of $\LUp_{k+1}$, have the property
       that $V_{k+1} \ge 2V_k-1$.}
    \figlab{leveldoubles}
\end{figure}

Note that the lemma also holds trivially when $V_k=1$, except that
then it only implies the trivial inequality $V_{k+1}\ge 1$.

\subsection{Upper Bounds}

We now complete the proofs of both \thmref{case1} and \thmref{case2}.

\paragraph{Proof of \thmref{case1} (Case (i)).}
By assumption, in this case $\LUp_0$ has at least two vertices.
Hence, $V_0\ge 2$, and \lemref{double} implies that\footnote{Every
   vertex of $\LUp_0$ is also a vertex of $\LUp_1$.} $V_1 \ge 3$, and
in general $V_k \ge 2^k+1$, as is easily verified, for every
$k\le k_0-1$, where $k_0$ is the index introduced prior to
\lemref{findv0}. Hence, since the number of (distinct) vertices of
$\A(L)$ is at most $\binom{n}{2}$, it follows that after at most
$2\log n - 1$ upper levels, the assumption of \lemref{double} can no
longer hold, and, at the next upper level, which we have denoted as
$k_0$, we get at least one vertex of $\LUp_{k_0-1}$ that lies strictly
above $\LUp_{k_0}$, and, by \lemref{findv0}, any such vertex has
maximum level (and only these vertices have this property). This
completes the proof of \thmref{case1}.  $\Box$

\medskip

\paragraph{Proof of \thmref{case2} (Case (ii)).}
This case is slightly more involved. Let $v$, $L_v$, $K$, $D$, $h$,
$D_0$ and $L_0$ be as defined prior to the theorem
statement. %
In this case, each of the first $D$ upper levels $\LUp_0$, $\LUp_1$,
\ldots, $\LUp_{D-1}$ will have just a single vertex, namely $v$, but
$\LUp_{D}$ has at least one new vertex that is an intersection of some
line of $K$ with either the $(D+1)$-st highest left ray or the
$(D+1)$-st highest right ray emanating from $v$ (rays are numbered
starting at 1).

From this level on, \lemref{double} can be applied, and it implies
that there exists a level %
among the %
subsequent levels %
$\LUp_D$, $\LUp_{D+1}$, \ldots, $\LUp_{D + 2 \log n}$ of $\A(L)$, for
which there exists a vertex that lies strictly above the level, and,
at the first time this happens, any such detached vertex has maximum
level in $\A(L)$, by \lemref{findv0} (and only these vertices have
this property). If $D\le 2\log n$ then no line is removed, and both
claims of the theorem (that all the maximum-level vertices of $\A(L)$
are vertices of $\A(L_0)$, and that the upper level in $\A(L_0)$ of
the maximum-level vertices is at most $4\log n$) hold; the first is
trivial and the second follows from %
$D + 2 \log n \le 4 \log n$. Assume then that $D > 2\log n$.  In this
case $D_0 = 2\log n$. %
Since no line $\ell \in L_v$ contributes to both prefixes of $L_v^-$
and $L_v^+$ of length $D$, at least $2D$ upper levels of $\A(L)$ pass
through $v$. In particular, $v$ lies on all levels $\LUp_0$ to
$\LUp_{D + 2 \log n}$.  We claim that none of the $2h$ lines removed
from $L$ can meet any of the upper levels
 $\LUp_h=\LUp_{D-2\log n}$ to
$\LUp_{D + 2 \log n}$
 of $\A(L)$, except for passing through it
at $v$.  Indeed, any line $\ell$ that contributes a ray to the top $h$
rays of $L_v^+$ passes to the right of $v$ above at least
$D_0 = 2\log n$ other lines of $L_v$, none of which has been removed,
so $\ell$ passes below all these lines to the left of $v$ and thus
cannot meet the topmost %
$D + 2\log n$ levels of $\A(L)$ to the left of $v$, and it clearly
cannot do so to the right of $v$. \figref{D-h-D0} illustrates this
argument. The argument for lines that contribute a ray to the top $h$
rays of $L_v^-$ is fully symmetric. %
We conclude that upper levels $\LUp_h=\LUp_{D-2\log n}$ to
$\LUp_{D + 2 \log n}$ of $\A(L)$ are identical to levels $\LUp_0$ to
$\LUp_{4 \log n}$ of $\A(L_0)$, and hence the upper level of any point
in these levels (except for $v$) with respect to $L$ is $h$ plus its
upper level with respect to $L_0$. %
Thus their upper level with respect to $L_0$ is at most
$D + 2 \log n - h = 4 \log n$. All this completes the proof of the
theorem.  $\Box$

\begin{figure}[h]
    \centerline{\includegraphics{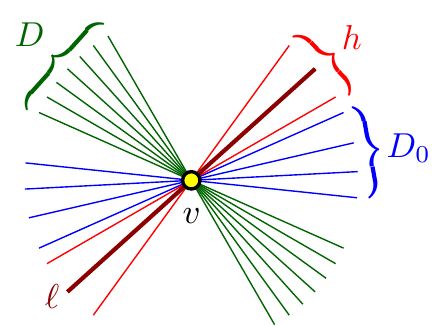}}
    \caption{\sf The prefixes of length $D$ of $L_v^-$ and $L_v^+$ are
       indicated in green and red/blue respectively. No line of $L_v$
       contributes to both prefixes. To the left of $v$ any of the $h$
       red lines has at least $D + D_0 = D + 2\log n$
       lines above it. }
    \figlab{D-h-D0}
\end{figure}

\subsection{Lower Bound}

In this subsection we give a construction that satisfies the property
of Case (i), for which the upper level of all the maximum-level
vertices is $\Omega(\log n)$.  We put $m=2^t$, for some integer $t$,
and construct the set $P$ of the $2m+1$ points
$p_{-m},\ldots,p_{-1},p_0,p_1,\ldots,p_m$ on the parabola
$\gamma:\; y=x^2$, where
\begin{align*}
  p_0 & = (0,0) , \\
  p_i & = (3^{i-1},3^{2(i-1)}) ,\qquad\text{for $i=1,\ldots,m$} \\
  p_{-i} & = (-3^{i-1},3^{2(i-1)}) ,\qquad\text{for $i=1,\ldots,m$} .
\end{align*}

For each $j=0,\ldots,t$, we construct a set $L_j$ of $s_j = 2^{j+1}$
`dyadic' lines. Concretely, for each $j$ we set
$L_j = L_j^-\cup L_j^+$, where the $r$\th line in $L_j^+$ connects the
points $p_{(r-1)2^{t-j}}$ and $p_{r2^{t-j}}$, for $r=1,\ldots,2^j$,
and the lines of $L_j^-$ are reflected copies of the lines of $L_j^+$
about the $y$-axis (so the $r$\th line in $L_j^-$ connects the points
$p_{-(r-1)2^{t-j}}$ and $p_{-r2^{t-j}}$, for $r=1,\ldots,2^j$). We put
$L := \bigcup_{j=0}^{t=1} L_j$, and note that
$|L| = \sum_{j=0}^t 2^{j+1} = 2^{t+2}-2$. See \figref{lower} for an
illustration.

\begin{figure}
    \centerline{\includegraphics{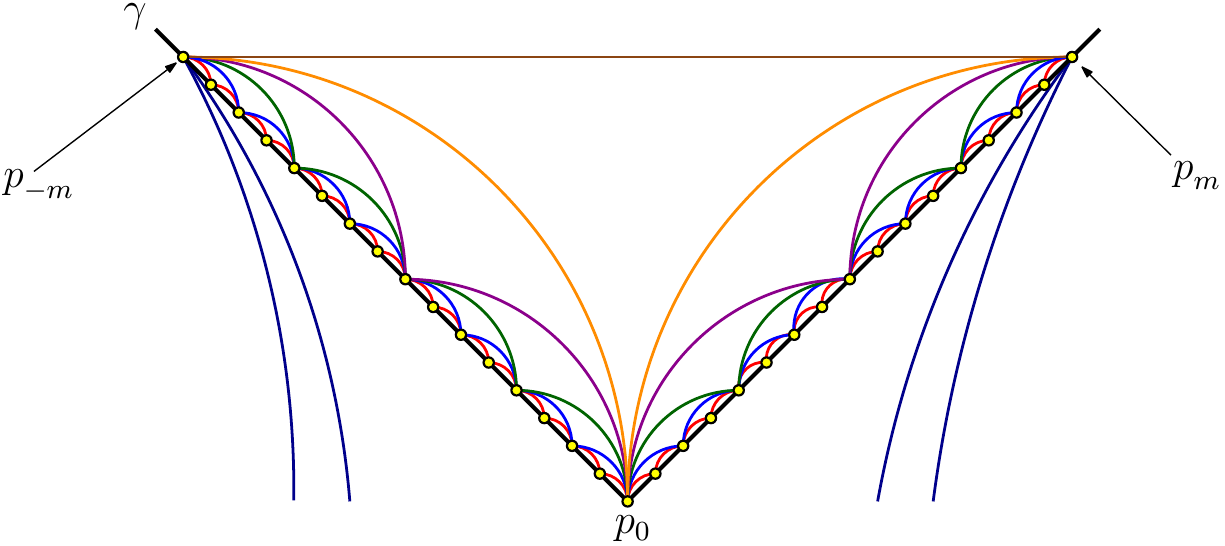}}
    \caption{\sf A schematic illustration of the construction, where
       the parabola is flattened to a V shape and the scale is
       logarithmic.}
    \figlab{lower}%
\end{figure}

\medskip
\begin{lemma} \lemlab{claim} (a) All the intersection points of the
    lines of $L$
    are either points of $P$ or lie below the parabola $\gamma$. \\
    (b) All these intersection points lie in the $x$-range between
    $p_{-m}$ and $p_m$.
\end{lemma}

\noindent{\bf Proof.}
Associate with each line $\ell\in L$ the arc $\gamma_\ell$ of $\gamma$
between the two points of $P$ that $\ell$ connects.  By construction,
each pair of these arcs are either openly disjoint or nested within
one another. This immediately implies (a). For (b), consider a pair of
lines $\ell, \ell'\in L$. The claim trivially holds when $\gamma_\ell$
and $\gamma_{\ell'}$ are openly disjoint, as the intersection point
lies in the $x$-range between the two arcs.  Assume then that the arcs
are nested, say $\ell$ connects $p_u$ and $p_v$, $\ell'$ connects
$p_w$ and $p_z$, and $u\le w < z \le v$. If $u=w$ or $z=v$, the lines
intersect at a point of $P$ and the claim follows, so assume that
$u<w<z<v$. The construction allows us to assume, without loss of
generality, that $0\le u<w<z<v$. Assume first that $u>0$. To simplify
the notation, write $a=3^{u-1}$, $b=3^{v-1}$, $c=3^{w-1}$, and
$d=3^{z-1}$. Let the intersection point be $(x,y)$.  Then we have
\begin{align*}
  \frac{y-a^2}{b^2-a^2} & = \frac{x-a}{b-a} ,
  & \text{for the line  passing through $(a,a^2), (b,b^2)$ and
    $(x,y)$} \\
  \frac{y-c^2}{d^2-c^2} & = \frac{x-c}{d-c} , 
  & \text{for the line  passing through $(c,c^2), (d,d^2)$ and
    $(x,y)$} \;, \\
\end{align*}
and it thus follows that
$$
x = \frac{ab-cd}{a+b-c-d} .
$$
We claim that $-b < x < b$, from which (b) follows. Observing that
$b > 3c$ and $b > 3d$, the denominator is positive, so we need to show
that
$$
-b(a+b-c-d) < ab - cd < b(a+b-c-d) .
$$
Divide everything by $a^2$, and put $C=c/a$, $D=d/a$, and $B=b/a$.  We
thus need to show that
$$
-B(1+B-C-D) < B-CD < B(1+B-C-D) .
$$
The right inequality becomes $(B-C)(B-D) > 0$, which clearly holds as
$B>C, D$.  The left inequality becomes $B^2 + 2B > CD + BC + BD$,
which also holds since $C, D \le B/3$.

The case $u=0$ is handled in exactly the same manner, except that we
replace $a$ by $0$. It is easily checked that the required
inequalities continue to hold. This completes the proof.  $\Box$

\medskip
To complete the construction, we generate two
additional %
arbitrary lines that pass through $p_m$ and are
contained in the acute-angled cone spanned by the tangent to $\gamma$
at $p_m$ and the vertical line through $p_m$, and apply the same
construction at $p_{-m}$. Altogether we obtain a set $L'$ of
$n = 2^{t+2} + 2$ lines. It is easily checked that any intersection
point formed by any of the new lines also lies in the $x$-range
between $p_{-m}$ and $p_m$. This, combined with \lemref{claim}, imply
that the upper level of any vertex of $\A(L')$ that lies below
$\gamma$ is at least $t+1$, implying that the actual level of any such
vertex is at most $n-t-3$. It thus remains to calculate the levels of
the points of $P$.

For $p_m$, we have $t+3$ lines passing through this point, and no line
of $L'$ passes above it, so its level is $n-t-3$.  The same holds for
$p_{-m}$. For any other $p_u$, with $u\ne 0$, let $j$ be the largest
integer such that $2^j$ divides $u$; for $u=0$ set $j=t$. Then, by
construction, there is exactly one line of $L_{i}$, for each
$i < t-j$, that passes above $p_u$, and two lines of $L_i$ are
incident to $p_u$, for each $i\ge t-j$. Hence the number of lines that
pass through or above $p_u$ is (exactly)
$$
2(j+1) + (t-j) = t+j+2 ,
$$
implying that the level of $p_u$ is $n -t - j - 2$.  The maximum value
is attained for $j=0$, which is $n-t-2$. This is therefore the maximum
level of a vertex of $\A(L)$, and all the vertices with $j=0$ (those
with odd indices) have $t-1 = \Theta(\log n)$ lines of $L$ passing
above them; that is, their upper level is $\Theta(\log n)$.

\section{Algorithms}%
\seclab{algorithms}

We now present an efficient, $O(n\log n)$-time algorithm for each of
the two cases.

\paragraph{Case (i).}
Here we need to construct the $k:=2\log n$ upper levels of $\A(L)$ and
report any detached vertex (or, for that matter, all detached
vertices) of maximum level. We use the algorithm of Everett
\etal~\cite{erk-96}, but we want to run it on a set of lines in
general position. For this, we perturb each line
$\ell_1,\ldots,\ell_n$ of $L$, using a special kind of symbolic
perturbation that uses only parallel shifts. That is, each line
$\ell_i$, with equation $y = a_ix+b_i$, is replaced by a line
$\ell'_i$, given by $y = a_ix+b_i+\eps_i$, where the $\eps_i$'s are
symbolic infinitesimal values, satisfying
$\eps_1 \gg \eps_2 \gg \cdots \gg \eps_n$.  Let $L'$ denote the set of
perturbed (actually, shifted) lines. We apply the algorithm of
\cite{erk-96} to $L'$, to compute the $k$ upper levels of $\A(L')$, in
time $O(nk+n\log n) = O(n\log n)$.

We resolve any comparison that the algorithm performs using the
varying orders of magnitude of the $\eps_i$'s. As a concrete
illustration, consider a comparison between (the $x$-coordinates of)
two intersection points of some line $\ell_i$ with two other lines
$\ell_j$, $\ell_m$.
(We may assume that $\ell_i$ is parallel to neither $\ell_j$ nor to
$\ell_m$.) %
The $x$-coordinates of the two intersection points are
$$
x_{i,j} = - \frac{b_j-b_i}{a_j-a_i} - \frac{\eps_j-\eps_i}{a_j-a_i} ,
\qquad\text{and}\qquad x_{i,m} = - \frac{b_m-b_i}{a_m-a_i} -
\frac{\eps_m-\eps_i}{a_m-a_i} .
$$
When comparing these values, if the non-infinitesimal terms in these
expressions are unequal, the outcome of the comparison is
straightforward.  If they are equal, the difference between these
$x$-coordinates is a linear combination of $\eps_i$, $\eps_j$, and
$\eps_m$. Using the different orders of magnitude of these parameters,
we can easily obtain the sign of the comparison.

Similar actions can be taken for any of the other basic operations
that the algorithm performs. Clearly, the cost of each basic
operation, including the cost of resolving comparisons via the
symbolic perturbation technique, is still constant.

It is straightforward to extract from the output of the algorithm the
top $k$ levels as a collection of edge-disjoint
$x$-monotone polygonal curves.

\paragraph{Transforming each perturbed level into the corresponding
   level in the original arrangement.}

Fix some index $j\leq k$.  We delete all the infinitesimal edges in
$\LUp_j$ of $\Arr(L')$ to obtain a left-to-right sequence
$s_1,s_2,\ldots,s_q$, where $s_1$ and $s_q$ are rays and the remaining
$s_i$'s are bounded segments.  The $x$-projections of these elements
are pairwise openly disjoint, and they might have (infinitesimal) gaps
between them (due to the deletion of in-between infinitesimal edges). We define
the function $F$ so that it associates with each segment $s_i$, which
is supported by some (unique) perturbed line $\hat{\ell}_m$, the
unperturbed $\ell_m$, namely $F(s_i)=\ell_m$.  With each pair of
consecutive segments $s_i, s_{i+1}$, we associate the intersection
point of their associated lines $F(s_i)\cap F(s_{i+1})$,
unless $F(s_i)=F(s_{i+1})$. In the latter case, the level progresses
from $s_i$ to $s_{i+1}$ along the same line $F(s_i)=F(s_{i+1})$ of
$L$, and we therefore merge the segments $s_i$ and $s_{i+1}$ into
a single segment, ignore the activity in the perturbed level near the
infinitesimally-separated endpoints of $s_1$ and $s_2$, and proceed to
handle the next pair $s_{i+1}, s_{i+2}$.  See
\figref{straight:through}.

\begin{figure}[h]
    \phantom{}\hfill%
    \includegraphics[page=1]{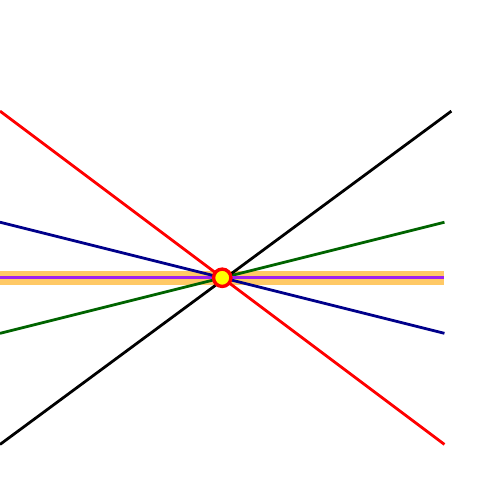}%
    \hfill%
    \includegraphics[page=2]{figs/no_turn_2}%
    \hfill%
    \includegraphics[page=3]{figs/no_turn_2}%
    \hfill%
    \phantom{}%
    \caption{\sf The situation in the unperturbed and the perturbed
       arrangements (in the left and center subfigures, respectively) near a vertex of some upper level $\LUp_j$ where the level
       does not bend. The thick segments in the right subfigure are
       infinitesimal, and they all collapse into the original single
       vertex (marked in the left subfigure). }
    \figlab{straight:through}%
\end{figure}

These intersection points are now the breakpoints of the level
$\LUp_j$ of $\A(L)$, which is a polygonal line with segments
connecting neighboring breakpoints, and each segment is contained in a
suitable line of $L$.  Finally, we complete $\LUp_j$ by adding the ray
portion of $F(s_1)$ from $F(s_1)\cap F(s_2)$ to the left, and the ray
portion of $F(s_q)$ from $F(s_{q-1})\cap F(s_q)$ to the right.

As is easily verified, this procedure yields the top $k+1$ levels of
$\A(L)$ (namely, the top levels $0,1,\ldots,k$).
This follows by observing that the level, as well as the upper level,
of each edge of non-infinitesimal length of $\A(L')$ is equal to the
level, or upper level, of the corresponding edge of $\A(L)$. Moreover,
the level and the upper level of any edge of non-infinitesimal length
(whether in $\A(L')$ or in $\A(L)$) add up to $n-1$, so either of
these two quantities determines the other one.

We note though that this is not true for vertices, where the level and
the upper level of a vertex can add up to any value between $n-2$ and
$0$.  To compute the level of a vertex $v$, we need to know both the
upper level of $v$ and its degree. While we know the upper level of
each vertex $v$ encountered in the construction, we may not know its
degree, as we might not have encountered all its incident lines. More
precisely, the algorithm of \cite{erk-96} does encounter only the
lines that are incident to $v$ and contribute edges that are adjacent
to $v$ and belong to the at-most-$k$ upper level; see a review of (our
version of) the algorithm in the appendix.  This is not an issue when
$v$ is an \emph{internal} vertex, that is, when $v$ lies strictly
above the $k$-upper level, as all its incident lines participate in
the $k$ top levels, but it may be problematic for vertices that lie on
the $k$\th level itself; see an example in \figref{neutral}.  Since we
know, by \lemref{findv0}, that all the maximum-level vertices are
internal (i.e., detached) vertices, for $k=2\log n$, the procedure will compute their
correct levels, and will let us find all the vertices of maximum
level.

\begin{figure}
    \centerline{\includegraphics{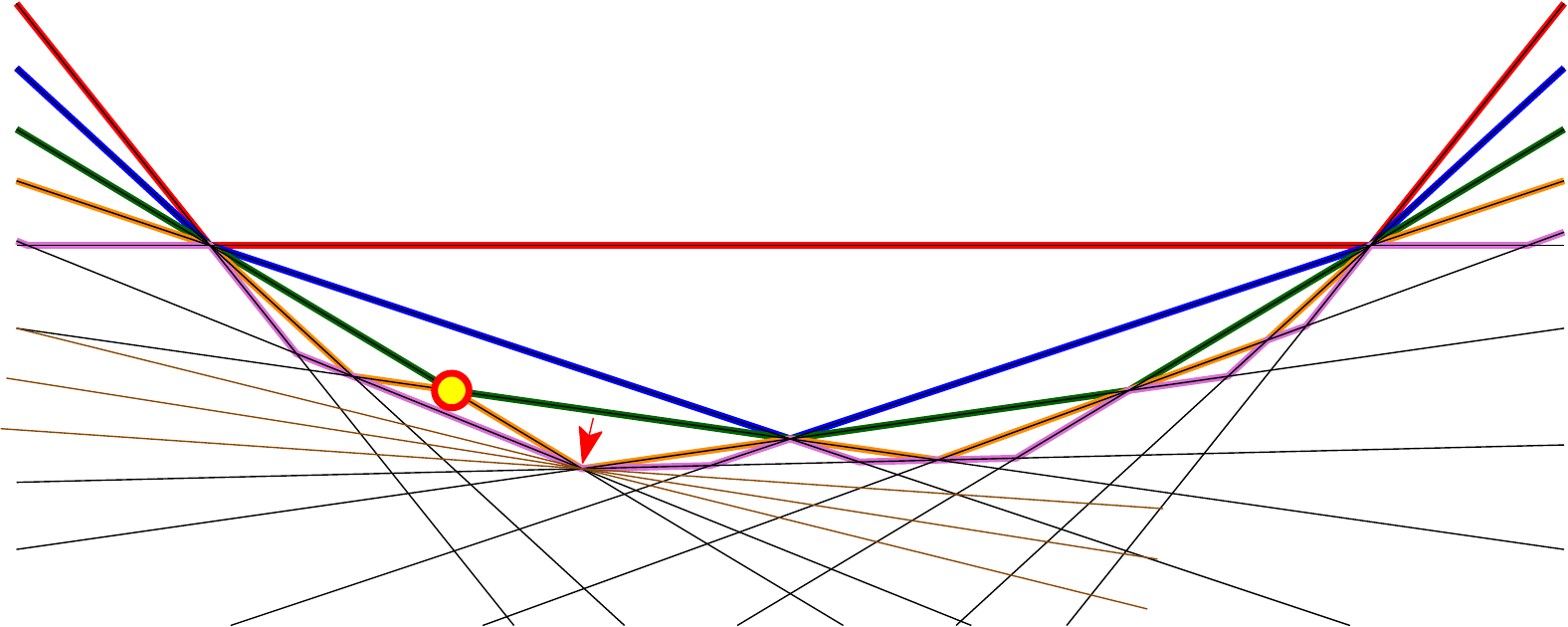}}
    \caption{\sf When computing the top $k$ levels, we might not know (the degree, and thus) the
       level of a vertex that is on the bottommost $k$-upper level,
       such as the arrow-marked vertex. We do know, though, the level of
       any vertex, like the circle-marked vertex, that lies strictly
       above the $k$-upper level.}    
    \figlab{neutral}%
\end{figure}

To recap, we have shown that in Case (i) we can find all the
maximum-level vertices in $O(n\log n)$ time.\footnote{%
   Notice that in the above description we do not aim to find the
   critical upper level $k_0$, and only rely on the property that the
   maximum-level vertices must be internal vertices of the
   at-most-$2\log n$ upper level. Thus the algorithm might also
   examine vertices that lie on or below the critical level.}

\paragraph{Case (ii).}
Here we first retrieve, in $O(n)$ time, the single vertex $v$ of the
upper envelope and the set $L_v$ of all its incident lines. We obtain
the corresponding sets $L_v^-$, $L_v^+$ of their left and right rays,
respectively, and sort each of them in descending order, as prescribed
earlier. We take the complementary set $K = L\setminus L_v$, compute
its upper envelope $E_K$, and test each ray of $L_v^-\cup L_v^+$ for
intersection with $E_K$. All this takes $O(n\log n)$ time, and yields
the parameter $D$.

We compute the parameters $h$, $D_0$, as defined in \secref{intro},
and remove from $L$ the $h$ lines that contribute the $h$ topmost rays
to $L_v^-$ and the $h$ lines that contribute the $h$ topmost rays to
$L_v^+$.  We then compute the at-most-$4 \log n$-upper level in the
arrangement $\A(L_0)$ of the set $L_0$ of the surviving lines, and
report all vertices of maximum level (in $\A(L_0)$), as we did in Case~(i).
We claim that these are also the maximum-level vertices in $\A(L)$.
Indeed, this follows from the construction, observing that (a) for any
such vertex $u$, other than $v$, the number of lines of $L$ that pass
above $u$ is exactly $h$ plus the number of lines of $L_0$ that pass
above $u$, (b) these upper levels do not contain any vertex of $\A(L)$
that is not a vertex of $\A(L_0)$, and (c) for any other point $u$
that lies below these upper levels, the number of lines of $L$ that
pass above $u$ is at least $h$ plus the number of lines of $L_0$ that
pass above $u$.

That is, we have shown that in Case (ii) too we can find all the
maximum-level vertices in $O(n\log n)$ time. In summary, we have
finally managed to solve Exercise 8.13~in \cite{bcko-08} for the case where all the input lines are distinct. That is, we
have:
\begin{theorem}
    All the maximum-level vertices in an arrangement of $n$ distinct
    lines in the plane can be computed in $O(n\log n)$ time.
\end{theorem}

\section{The case of coinciding lines}%
\seclab{coincide}

We now turn to the more degenerate setup where the lines of $L$ can
repeat themselves.  Let $\Gamma=\{\gamma_1,\ldots,\gamma_m\}$ be the
set obtained from $L$ by removing duplicates. The lines of $\Gamma$
are pairwise distinct, and we denote by $f$ the function that maps
each line in $L$ to its representative (overlapping) line in $\Gamma$.
For each $\gamma\in\Gamma$ we denote by $\mu(\gamma)$ its
\emph{multiplicity}, namely the number of lines $\ell\in L$ satisfying
$f(\ell) = \gamma$. We naturally have
$\sum_{\gamma\in\Gamma} \mu(\gamma) = n$.

The level $\lambda_\Gamma(p)$ of a point $p$ in $\A(\Gamma)$ is
defined, as before, to be the number of lines of $\Gamma$ that pass
strictly below $p$.  The situation is somewhat different for $\A(L)$.
For any point $p$ in the plane define
\begin{equation*}
    S(p) := \sum_{\gamma \in \Gamma\,:\,\gamma \text{ passes
          below }p} \mu(\gamma).
\end{equation*}%
If $p$ is a vertex of $\A(L)$ then its level in $\A(L)$  is $\lambda_L(p) =
S(p)$. If $p$ lies in the relative interior of an edge of $\A(L)$
then it lies on some line $\gamma$ of $\A(\Gamma)$, and we say that
\emph{$p$ lies at level $k$} in $\A(L)$ if
\begin{equation} \eqlab{deglevel} S(p) \le k < S(p) + \mu(\gamma). %
\end{equation}
In words, an edge $e$ of $\A(L)$ (that is, of $\A(\Gamma)$) may
participate in several consecutive levels, depending on its
multiplicity. This extends to edges a similar phenomenon (already noted) that holds
only for vertices in arrangements of distinct lines. 

\newcommand{\figX}[1]{
   \begin{minipage}{0.22\linewidth}
       \smallskip%
       
       \includegraphics[page=#1,width=0.99\linewidth]{figs/w_level}%
   \end{minipage}
}

\begin{figure}[h]
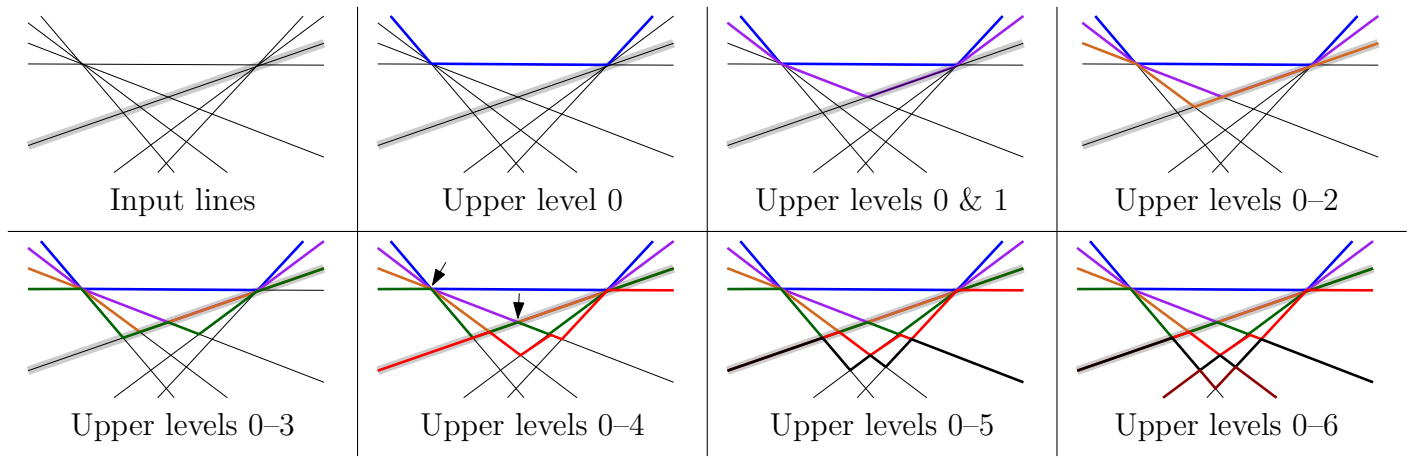

    \begin{tabular}{c|c|c|c}
      \figX{1}
      &
        \figX{2}
      &
        \figX{3}
      &
        \figX{4}\\
      Input lines
      & Upper level 0
      & Upper levels 0 \& 1
      & Upper levels 0--2 $\Bigl.$\\
      \hline
      \figX{5}
      &
        \figX{6}
      &
        \figX{7}
      &
        \figX{8}
      \\
      Upper levels 0--3
      &
        Upper levels 0--4
      &
        Upper levels 0--5
      &
        Upper levels 0--6 $\Bigl.$
      \\      
    \end{tabular}
    \caption{\sf The thick line has multiplicity $2$, all other lines
       have multiplicity $1$. Upper level 4 is the first upper level that
       has detached vertices above it, marked by arrows in the sixth
       subfigure. }
    \figlab{coin:level}%
\end{figure}

The $k$-level $\LDown_k$
in $\A(L)$ is the closure of the union of all edges $e$ of
$\A(\Gamma)$ that lie at level $k$ (in $\A(L)$, according to the
definition in \Eqref{deglevel}).  Fully symmetric definitions apply to
the upper level. 
See \figref{coin:level} for an illustration.
Note that, as in the case of distinct lines (and
even more so in this setup), the level does not necessarily turn at
every vertex $v$ that it reaches: it could pass through $v$ staying on
the same line of $\Gamma$; see for example upper levels 2, 3 and 4 in
\figref{coin:level} for an illustration.
Note also that in this setup different levels may share edges of $\A(\Gamma)$.

As in the case of distinct lines, we wish to find the smallest upper
level $k_0$ in $\A(L)$ for which there is a vertex in $\A(L)$ that
lies strictly above $\LUp_{k_0}$. All these (detached) vertices will be our
desired maximum-level vertices, a property that is established
rigorously in the following lemma.
\begin{lemma}%
    \label{detached:coin}%
    Let $k_0$ be the first index for which $\A(L)$ contains a vertex
    that lies strictly above the $k_0$-upper level $\LUp_{k_0}$ of
    $\A(L)$.  Then all these `detached' vertices (and only those) are the maximum-level
    vertices of $\A(L)$.
\end{lemma}
\noindent{\bf Proof.}
Let $v$ be one of these detached vertices. We have
$\lambda_L(v) = \mu(e) + S(e)$, where $e$ is the edge of $\LUp_{k_0}$
within $\A(\Gamma)$ lying vertically below $v$, and $\mu(e)$ (resp.,
$S(e)$) is the value $\mu(p)$ (resp., $S(p)$) for any point $p\in e$.
If $v$ lies vertically above a vertex of $\LUp_{k_0}$, %
apply this definition to an edge $e$
of $\LUp_{k_0}$ incident to this vertex.
By definition, and since $k_0$ is the smallest upper level with this
property, we have $\lambda_L(v) = n-k_0$. On the other hand, let
$w$ be any vertex lying on or below $\LUp_{k_0}$. Assume for
simplicity that $w$ is a vertex of $\LUp_{k_0}$. Move, as before, from
$w$ to a point $w'$ slightly to the left of $w$ along the line of $L$
that lies on $\LUp_{k_0}$ just to the left of $w$. Any line (of $L$)
that passes strictly below $w$ also passes below $w'$, so, again by
definition, $\lambda_L(w) \le \lambda_L(w') < \mu(e) + S(e) = n-k_0$, where $e$ is the edge of $\A(L)$ (or rather of $\A(\Gamma)$) that contains $w'$.
The same argument applies to vertices $w$ below $\LUp_{k_0}$; the
level $\lambda_L(w)$ can only get smaller.  $\Box$

\medskip

Due to the non-standard definition of levels in $\A(L)$, it seems
difficult (and at the moment we do not know how) to apply the method of the previous sections to the current
setting. Instead we proceed as follows. We first perturb the lines in
$L$ to obtain a set of lines $\hat{L}$, which induces a
degeneracy-free arrangement $\A(\hat{L})$. We then work in tandem with
both this perturbed arrangement, and the arrangement $\A(\Gamma)$. We
use the arrangement $\A(\hat{L})$ to carry out a binary search on its
upper levels. Each time we extract a specific $k$-upper level from
$\A(\hat{L})$, we transform it into a polygonal curve $\pi_k$, which
is contained in the union of the lines of $\Gamma$, and \emph{which is precisely the $k$-upper level of  $\A(L)$, as defined above}. We look for the
smallest $k$ for which there is at least one vertex in $\A(\Gamma)$
strictly above $\pi_k$. In the remainder of this section we describe
the perturbation of the lines of $L$ into those of $\hat{L}$, how we
carry out the binary search over the upper levels of $\A(\hat{L})$,
and how we detect whether, for a given $k$, there is a vertex of
$\A(\Gamma)$ above $\pi_k$.

\paragraph{The perturbation.} We apply symbolic perturbation to the lines in $L$, using the parallel
shifting mechanism described in \secref{algorithms}, to obtain the set
$\hat{L}=\{\hat{\ell}_1,\ldots,\hat{\ell}_n\}$.  Notice that this
turns each line $\gamma\in\Gamma$ into $\mu(\gamma)$ parallel lines,
infinitesimally close to one another. We define another function $F$,
which maps each perturbed line $\hat{\ell}_i$ to the line
$\gamma_j\in \Gamma$ that overlaps with the original line $\ell_i$
whose perturbed counterpart is $\hat{\ell}_i$, namely
$F(\hat{\ell}_i)=f(\ell_i)$.

\vspace{1ex} Notice that, under the standard conventions about
symbolic perturbation, the arrangement $\A(\hat{L})$ is in general
position (except for lines overlapping the same $\gamma\in\Gamma$
being parallel to one another). We compute the $k$-upper level
$\LUp_k$ of $\A(\hat{L})$, using a standard procedure for this task
(see~\cite{DBLP:journals/siamcomp/EdelsbrunnerW86} and the appendix),
and then transform it into the aforementioned unbounded $x$-monotone polygonal curve
$\pi_k$, comprising non-infinitesimal portions (segments and rays) of the lines in
$\Gamma$, joined together at the infinitesimal gaps between them (when such gaps exist); see \figref{zig_zag}.
This is done exactly as in the procedure in \secref{algorithms} for
extracting the unperturbed level in degenerate arrangements that have
no coinciding lines.

\begin{figure}[t]
	\centerline{%
		\begin{tabular}{*{4}{c|}c}
			\includegraphics[page=1]{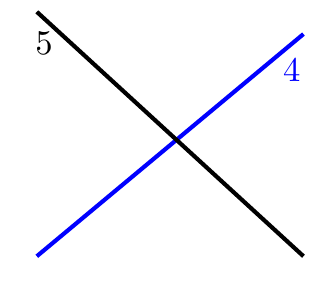}%
			&%
			\includegraphics[page=3]{figs/zig_zag}%
			&%
			\includegraphics[page=5]{figs/zig_zag}%
			&
			\includegraphics[page=4]{figs/zig_zag}%
			&
			\includegraphics[page=6]{figs/zig_zag}%
			\\
			(a)&(b)&(c)&(d)&(e)
		\end{tabular}%
	}
	\caption{\sf The construction of $\pi_k$ near a vertex, for $k=3$. %
		(a) The input lines with their multiplicities. %
		(b) The perturbed lines, and the level in the perturbed
		arrangement. %
		(c) The path $\pi_k$ in $\A(\Gamma)$ -- six infinitesimal edges have been
		deleted.  %
		(d) Another example of a level,now for $k=4$ and
		(e) its resulting path $\pi_k$ in the arrangement $\A(\Gamma)$.
	}
\figlab{zig_zag}
\end{figure}

\paragraph{The binary search for computing $k_0$.}
To compute $k_0$ and the set $V_0$ of the detached vertices, we
perform a binary search over the upper levels in $\A(\hat{L})$ in the
following manner.  Initially the range of potential levels is $[1,n]$
and we set $k$ to be $\lfloor n/2 \rfloor$.  We compute the $k$-upper
level $\LUp_k$ of $\A(\hat{L})$ (see below for details), and transform
$\LUp_k$ into $\pi_k$ as described above. We then compute the portions
of the lines in $\Gamma$ that lie above $\pi_k$ (see details below).
Again, this is a collection of line segments and rays, which we 
denote by $\Delta_k$. We now need to determine whether any pair of
elements of $\Delta_k$ intersect strictly above $\pi_k$ (i.e., they intersect at their relative interiors), which we can
do using the decision procedure to be described below. If there is no
such intersection, then the current $k$ is too small, and the new
range is the bottom half of the current range, otherwise we set the
new range to be the top half. We set $k$ to be the middle index of the
new range and recurse. It may be the case that we do not find a
desired level with a vertex above it, in which case the maximum level
of any vertex of $\A(\Gamma)$ is zero; this can only happen if all the
lines meet in a single point, which is the single vertex of the
arrangement.

To complete the description of the algorithm, we detail two procedures, which will be applied at each step of the binary search, for: (i) finding the set $\Delta_k$ of segments and rays that lie above $\pi_k$, and (ii) deciding whether the curves in $\Delta_k$ intersect above $\pi_k$. Also, we describe how to find the set of vertices $V_0$, once the level $k_0$ had been determined. 

\paragraph{Computing the set $\Delta_k$.}    
In order to determine whether there is a vertex of the arrangement  $\A(L)$ above $\pi_k$, we first need to
collect the
portions of lines in $\Gamma$ that lie above $\pi_k$, To do so, we find the
leftmost vertex of the arrangement $\Arr(\Gamma)$ in $O(n \log n)$
time and project it vertically onto $\pi_k$. We then add a breakpoint
$b_L$ along $\pi_k$ slightly to the left of this projection point and
substitute the portion of the ray of $\pi_k$ emanating from $b_L$ to
the left by the upward vertical ray from $b_L$. We apply a symmetric
modification at the rightmost vertex of $\Arr(\Gamma)$, and replace the
right ray of $\pi_k$ with the segment connecting the rightmost vertex
of $\pi_k$ with the new point $b_R$ along $\pi_k$ and an upward
vertical ray from $b_R$.

Denote this modified version of $\pi_k$ by $\pi'_k$. We now compute
the set $\Delta_k$ of line segments comprising all the portions of
lines of $\Gamma$ that lie above $\pi'_k$, each represented by its
left and right endpoints. (Notice that, since we use the modified version $\pi'_k$, the set $\Delta_k$ contains segments only, and no rays.)  We intersect the lines in $\Gamma$ with the
upward vertical ray from $b_L$, to obtain some of the left endpoints
of segments in $\Delta_k$ (which are in fact internal points on the corresponding original rays). We store these endpoints in an array $W$,
which has an entry (not always occupied) for every line in $\Gamma$. Initially we set $W[\ell]$:=null for every line $\ell\in\Gamma$.  Additional endpoints
are detected by moving along $\pi'_k$ from left to right and carefully examining, for each vertex $v$ of the original $\pi_k$, the set $\Gamma(v)$ of all the lines of $\Gamma$ that are incident to $v$.

 To determine $\Gamma(v)$, we consider the (one or two) lines that contain the edges of $\pi_k$ incident to $v$, together with all the infinitesimal edges
that have been produced as part of $\LUp_k$ within $\Arr(\hat{L})$, and have been collapsed to $v$.
Consider such an infinitesimal edge $e$. Let
$\hat{\ell}_e$ be the perturbed line containing $e$, and let $v$
be the vertex of $\pi_k$ to which $e$ will be contracted during the
process of constructing $\pi_k$ (which may in particular unite two
collinear segments into a common segment). The line $f(\hat{\ell}_e)$
is split by $v$ into a leftward and a rightward ray. Consider the
leftward ray, and compare its slope with that of the line supporting
the edge $g$ immediately to the left of $v_k(e)$ along $\pi_k$. If the
ray has a smaller slope than $g$, then $v_k(e)$ is the right endpoint
of a segment whose left endpoint is stored in $W$. We add this segment
to $\Delta_k$ and remove the corresponding entry from $W$. For the
rightward ray we compare its slope with the slope of the line containing the edge $h$
along $\pi_k$ immediately to the right of $v_k(e)$. If it has a larger
slope than the line containing $h$, then we insert $v_k(e)$ into $W$ at the entry for
$f(\hat{\ell}_e)$, as this is the left endpoint of a segment that will
eventually be added to $\Delta_k$. (Notice that $f(\hat{\ell}_e)$ may contribute to $\Delta_k$ two segments incident to $v$.) Finally we intersect the upward
vertical ray from $b_R$ with each of the lines in $\Gamma$ and using
$W$ we form the corresponding segments (representing right rays) and add them to
$\Delta_k$. 

Since we are using the infinitesimal edges of $\Arr(\hat{L})$, we may encounter a segment of $\Arr(\Gamma)$ that should be added to $\Delta_k$ several times (as many times as its multiplicity)). We wish to report each such segment only once.  To do so, for any line $\ell$ of $\Gamma$ we only insert a left endpoint to $W[\ell]$ if this entry is null, namely it does not currently contain a left endpoint (if it already contains a left endpoint, this means that the left endpoint of this specific segment has already been detected due to another copy of $\ell$ in $\hat{L}$). Similarly, when we detect a right endpoint of a segment, we only report the segment if $W[\ell]$ contains a left endpoint---in that case we add the segment having these endpoints (the left endpoint in $W[\ell]$ and the corresponding right endpoint that we have just detected) to $\Delta_k$ and set  $W[\ell]$ to null.

This process of constructing the set $\Delta_k$ takes time
proportional to the complexity of the \emph{weighted $k$\th level} of
$\Arr( \hat{L} )$, where each vertex of the level is counted as many
times as there are lines passing through it. We show in
\lemref{k:level:weighted} in the next section that this quantity is
bounded by $O(nk^{1/3})$. This also bounds the size $|\Delta_k|$ of
$\Delta_k$.

\medskip

\paragraph{Deciding whether there is a vertex of the arrangement
	$\A(\Gamma)$ strictly above $\pi_k$.}
     We run a sweep-line algorithm over the segments in $\Delta_k$, to
     detect the first intersection that does not lie on $\pi_k$.
     Notice that all the vertices of $\pi_k$ are inserted into the
     event queue before the sweep starts.
     Such vertices occur at common endpoints of the segments, and are not intersections that we seek (which only occur within the relative interior of the segments).
     The same holds for the intersection of lines in $\Gamma$ with either $b_L$ or $b_R$---we insert them to the queue before the sweep starts and neither set contains a relevant vertex of the type we are looking for.

\paragraph{Finding the set $V_0$ of detached vertices.}
     After terminating the binary search at some index $k_0$, we need
     to find the set $V_0$ of \emph{all} detached vertices above
     $\pi_{k_0}$.
     We consider the set $\Delta_{k_0}$ of segments, and observe that
     all the vertices in $V_0$ are vertices of the lower envelope of
     $\Delta_{k_0}$.  Indeed, no segment of $\Delta_{k_0}$ can lie
     below any vertex $v$ of $V_0$, for then $v$ would be detached
     from an upper level with a smaller index.  We thus need to
     compute the lower envelope, which we can do using a standard
     divide-and-conquer technique (see, e.g., \cite{sa-95}). Since
     $|\Delta_{k_0}| = O(nk_0^{1/3})$, this construction takes
     $O(nk_0^{1/3}\alpha(n)\log n)$ time. 
     We output those vertices of the envelope that lie in the relative interiors of their incident segments (ignoring segment endpoints).

\paragraph{The overall complexity.}
     Computing the $k$-upper-level in $\A(\hat{L})$ takes
$O(nk^{1/3}\log^{2} k)$
time~\cite{DBLP:journals/siamcomp/EdelsbrunnerW86} (see also the
appendix). This time dominates the time of the other procedures carried out in a single step of the binary search. Hence, multiplying this by the number $O(\log n)$ of binary search
     steps, we thus conclude:
     \begin{theorem}%
         \label{main:coincide}%
         The maximum-level vertices in an arrangement of $n$ lines,
         where some lines may coincide, can be computed in
         $O(n^{4/3}\log^{3}n)$ time.
     \end{theorem}

\medskip
\noindent{\bf Remark.}
We can modify the binary search so that it first runs an exponential search from the top of the arrangement, and only reverts to standard binary search at the first time when the current level exceeds $k_0$. This improves the running time to $O(nk_0^{1/3}{\rm polylog}\;n)$, when $k_0 \ll n$. Obtaining such a sharp bound on $k_0$, or giving a construction in which $k_0 = \Theta(n)$, remains one of the open problems raised by the present work.

     \section{The complexity of the weighted $k$-level in degenerate
        arrangements}
     \seclab{weighted}

     Finally, we consider a related combinatorial question for
     degenerate arrangements.
     The resulting combinatorial bound, stated in
     \lemref{k:level:weighted}, has been used in the analysis of the
     previous section.

     As before, let $L$ be a set of $n$ lines, not necessarily in
     general position: we allow many lines to intersect in a single
     point, but assume that all the lines are distinct. Recall that
     the vertices of the $k$\th level $\LDown_k$ are not necessarily
     at level $k$. As a matter of fact, as already noted, if the
     degree of a vertex %
     $v$ of $\A(L)$ is $d$ and $k$ lines pass below $v$, then $v$
     belongs to the $d$ consecutive levels $k,k+1,\ldots,k+d-1$ of
     $\A(L)$.  Let $|\LDown_k|$ denote the complexity of $\LDown_k$,
     that is, the number of its vertices, and let $\omega(\LDown_k)$
     denote the \emph{weighted complexity} of $\LDown_k$, defined as
     the sum of the degrees of the vertices of $\LDown_k$. It is known
     \cite{d-98} that $|{\LDown_k}| = O(nk^{1/3})$ in the
     non-degenerate case (for this case we have
     $\omega(\LDown_k) = 2|\LDown_k|$). We strengthen this result for
     the degenerate case in the following lemma.

     \begin{lemma}
         \lemlab{k:level:weighted}%
         Let $L$ be a set of $n$ distinct lines in the plane, not
         necessarily in general position. Then
         $\omega(\LDown_k) = O(nk^{1/3})$.
     \end{lemma}

     \noindent{\bf Proof.}
     We convert the original arrangement of lines into an arrangement
     of pseudo-lines in general position, by making local changes in
     the vicinity of every vertex of degree greater than
     two. Furthermore, we ensure that, in the new arrangement, when
     the $k$\th level passes through the vicinity of any original
     vertex $v$ (so $v$ is a vertex of the original level), it visits all the pseudo-lines whose original lines
     pass through $v$, each along some segment thereof, before leaving this neighborhood.

     Consider such an original vertex $v$, of some degree
     $d=d(v)\ge 3$ (vertices of degree two require no action); see
     \figref{dense:vertex}(a).  The $k$\th level enters this vertex
     from the left, say on a line $\ell_L$, and leaves to the right,
     say on a line $\ell_R$. Assume that $\LDown_k$ forms a right turn
     at $v$ (the left turn case is handled in a similar fashion to
     what is described below, and it may also be the case that there
     is no turn, and the level enters and leaves $v$ along the same
     line).  A line that reaches $v$ from the left below the level,
     and leaves $v$ to the right above the level, is called
     \emphi{ascending}, a line that reaches $v$ from the left above
     the level but leaves $v$ to the right below the level is called
     \emphi{descending}, and a line that does neither is called
     \emphi{neutral}; such lines stay on the same side of the level
     both to the left and to the right of $v$.  In particular,
     $\ell_L$ and $\ell_R$ are neutral.
     Under the right-turn assumption, all the neutral lines pass above
     or on the level, both to the left and to the right of $v$; see
     \figref{dense:vertex}(a).

     We deform the batch of ascending lines into the kink-like
     structure $K_a$, and the batch of descending lines into the
     kink-like structure $K_d$, as depicted in
     \figref{dense:vertex}(b). We make the two middle portions of the
     kinks cross one another to the left of $v$, and below the (still
     untouched) batch of neutral lines. The lines of each class remain
     pairwise disjoint in a suitable small neighborhood $\Omega$ of
     the crossing, but we make every pair of them cross in some other
     portion of the respective kink, to the right of $\Omega$ and away
     from the lines of the other two classes.

     In addition, we deform the neutral lines within another small
     neighborhood $\Omega'$ of $v$ that is disjoint from any ascending
     or descending line (and from $\Omega$), so that each of them
     contributes an arc (of nonzero length) to their lower envelope
     within $\Omega'$.

     The construction ensures that the $k$\th level in the modified
     scenario proceeds along $\ell_L$ until it reaches $K_d$, then
     turns right along the first (leftmost) descending line, reaches
     $\Omega$, traces a zigzag pattern, alternating between ascending
     lines and descending lines, leaves $\Omega$ along the rightmost
     ascending line (this follows since the number of ascending lines
     is equal to the number of descending lines), reaches $\ell_L$
     again, and then proceeds along $\ell_L$ until it enters
     $\Omega'$; see the left magnifying glass in
     \figref{dense:vertex}(b). The deformation within $\Omega'$
     ensures that the level traces the lower envelope of the neutral
     lines, and leaves $\Omega'$ along $\ell_R$; see the right
     magnifying glass in \figref{dense:vertex}(b).

\begin{figure}[h]
    \phantom{}\hfill%
    \includegraphics[height=12.5em]{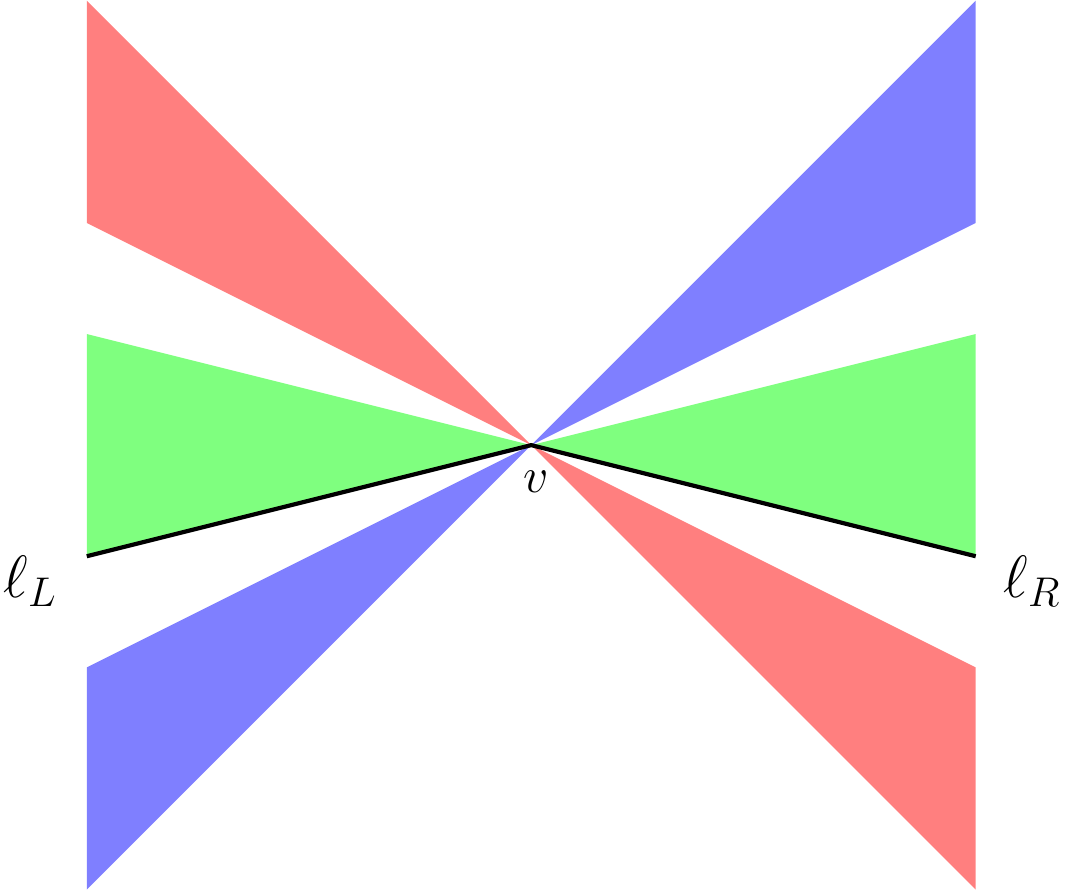} \hfill%
    \includegraphics[height=12.5em]{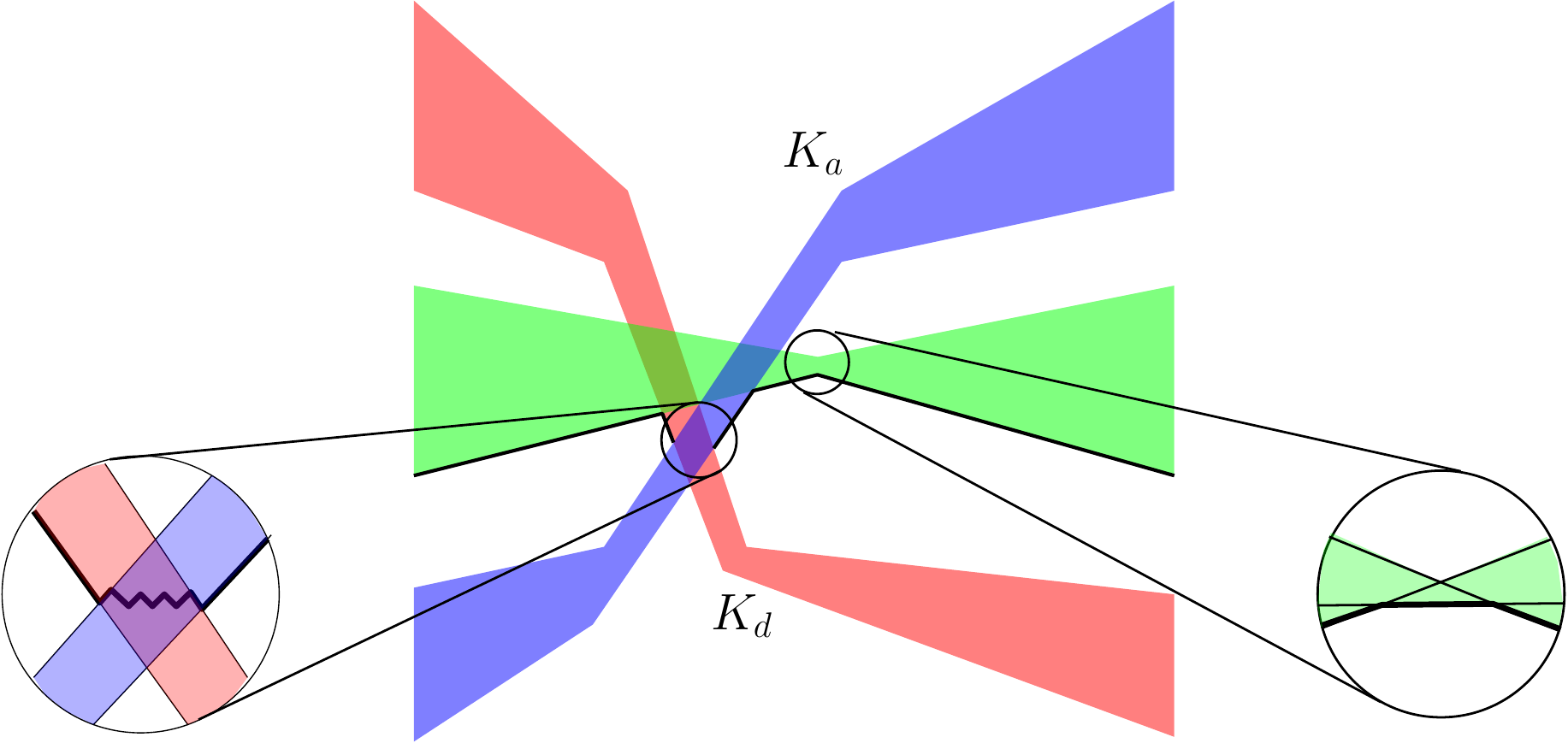}%
    \hfill \phantom{}
    
    \phantom{}\hfill \hspace{-5em} (a) \hfill \hspace{4em} (b) \hfill
    \phantom{}
    \caption{\sf (a) The lines passing through the original vertex
       $v$, with the $k$\th level marked in black. The ascending lines
       are marked in blue, the descending lines in red, and the
       neutral lines in green.  (b) The vicinity of the vertex $v$
       after the local transformation of the lines into pseudo-lines, where the neighborhoods $\Omega$ (to the left) and $\Omega'$ (to the right) are magnified.}
    \figlab{dense:vertex}
\end{figure}

The above transformation can be performed by deforming the lines
incident to $v$ only within an arbitrarily small square around $v$,
disjoint from all other vertices and their surrounding squares, so
that the new curves coincide with the original lines on the boundary
of and outside this square.  By construction, inside this square every
pair of modified curves intersect at exactly one point, and none of
these pairs intersect outside the square (even after the local perturbations taking place at square neighborhoods of other vertices). Hence the curves that come
from the original lines that are incident to $v$ constitute a family
of pseudo-lines.  We repeat this deformation for every vertex $v$ of
the $k$\th level of degree greater than $2$. For vertices $v$ that are
not on the level, whose degree is greater than $2$, a simpler
deformation suffices, only ensuring that each pair of lines that are
incident to $v$ intersect now, after their perturbations, at a distinct point, within a sufficiently
small neighborhood of $v$. All this results in a collection of $n$
pseudo-lines in general position, so that, for every vertex $v$ of
$\LDown_k$, each line incident to $v$ now contributes at least one
edge to the $k$\th level of the modified arrangement, within the
square corresponding to (and surrounding) $v$.

We have thus constructed an arrangement of pseudo-lines so that the
complexity of its $k$\th level is at least proportional to
$\omega(\LDown_k)$.  By the result of Tamaki and
Tokuyama~\cite{tt-03}, the complexity of the $k$\th level in an
arrangement of $n$ pseudo-lines is $O(nk^{1/3})$. This completes the
proof.  $\Box$

\paragraph{Acknowledgments.}  %
The authors thank Michal Kleinbort and Shahar Shamai for pointing out
the difficulty of the problem of finding the maximum-level vertex.

Work by Dan Halperin has been supported in part by the Israel Science
Foundation (grants no.~825/15 and~1736/19), by the Blavatnik Computer
Science Research Fund, and by grants from Yandex and from Facebook.

Work by Sariel Har-Peled was supported by an NSF AF award CCF-1907400.

Work by Micha Sharir has been supported in part by Grant 260/18 from
the Israel Science Foundation, by Grant G-1367-407.6/2016 from the
German-Israeli Foundation for Scientific Research and Development, and
by the Blavatnik Computer Science Research Fund.
%
%
%



%
\appendix

\section{A review of a variant of the algorithm of Everett \etal}
\apndlab{at:most:k}

In this appendix we present a variant of the algorithm by Everett
\etal~\cite{erk-96} for constructing the top $k$ levels of an
arrangement of lines.

\begin{theorem}{\rm (Based on Everett \etal~\cite{erk-96})}
    Given a set $L$ of $n$ lines in general position in the plane, and
    a parameter $k$, one can compute the top $k$ levels of $\A(L)$ in
    $O(n \log n + nk)$ time.
\end{theorem}
 
\noindent{\bf Proof.}      
We proceed in four steps. First, we discuss the case where all the
lines of $L$ show up on the upper envelope and derive a point location
data structure that we need in the other steps. In the second step, we
compute $k$ sets of lines $L'_1, \ldots, L'_k$ such that only lines in
$L' := L'_1\cup\cdots\cup L'_k$ appear in the $k$ top levels of
$\A(L)$.  Next we compute the $k$\th upper level of $\A(L')$, making
use of the decomposition computed in step 2 and the data structure
derived in step 1. Finally, we compute the part of the arrangement of
$\A(L')$ lying on or above the $k$\th upper level.

Let $L$ be a set of $n$ lines in the plane in general position,
meaning that no point is incident to more than two lines of $L$ ($L$
may contain parallel lines).  Consider the special case where all the
lines of $L$ show up on the upper envelope $E$ of $L$. Then $\A(L)$
has a special structure: except for the top face, which is bounded by
all $n$ lines, and the bottom face and the two unbounded faces
adjacent to the top face, which are wedges bounded by only two lines,
every other face is either a triangle or a quadrangle. The triangles
are all the other unbounded faces and all the other faces adjacent to
the top face, and the quadrangles are all the other faces. 
See \figref{boring}(left).

Point location in this arrangement is simple. We compute $E$, in
$O(n\log n)$ time (this amounts, in the special case under
consideration, to just sorting the lines of $L$ by their
slopes). Then, given a query point $q$ below $E$, we can compute the
face of $\A(L)$ containing $q$ in $O(\log n)$ time.  The simplest way
of doing this is to compute the (at most) two tangents from $q$ to
$E$, and use only the (at most four) lines incident to the points of
tangency to compute the desired face.
See \figref{boring}(right).

\begin{figure}[h]
    \hfill%
    \includegraphics[page=1]{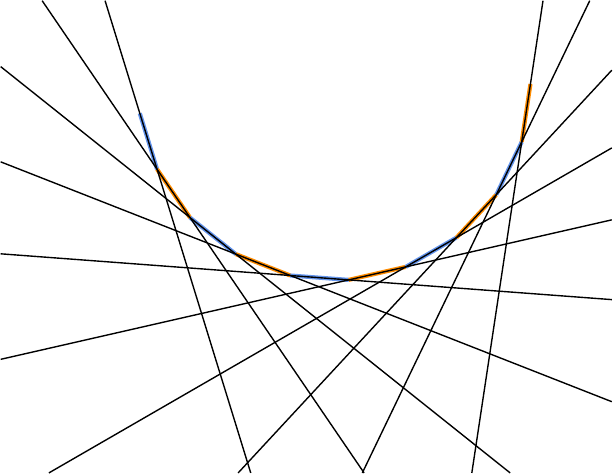}%
    \hfill%
    \includegraphics[page=2]{figs/boring}%
    \hfill%
    \phantom{}%
    \caption{\sf The special structure of the arrangement of lines
       that are in ``convex'' position, meaning that they all show up
       on their upper envelope. Left: The arrangement. Right: Locating a point below the envelope.}
    \figlab{boring}
\end{figure}

Consider now the general case, where we are given an arbitrary set $L$
of $n$ lines in general position, and a parameter $k$, and we want to
construct the $k$ top levels of $\A(L)$.  We apply the following
iterative `peeling' process to $L$, to obtain a sequence
$L_1,L_2,\ldots,L_k$ of subsets of $L$.  We set $L_1 = L$ and, for
each $i\ge 1$, we obtain $L_{i+1}$ from $L_i$ by constructing the
upper envelope $E_i$ of $\A(L_i)$, defining $L'_i$ to consist of all the
lines that show up on the envelope, and setting
$L_{i+1}:=L_i\setminus L'_i$.  A naive implementation of this process
takes $O(k\cdot n\log n) = O(nk\log n)$ time, but we can improve it to
$O(nk+n\log n)$ by noting that, once the lines of $L$ are sorted by
slope, we can compute the upper envelope (of any prescribed subset of
$L$) in linear time, e.g., by a dual version of Graham's scan
algorithm for computing convex hulls (see, e.g., \cite{bcko-08}). Set
$L' := L'_1\cup\cdots\cup L'_k$.  By construction, only the lines of $L'$
appear in (i.e., support the edges of) the $k$ top levels of $\A(L)$.

In the next step, we construct the $k$\th upper level of $\A(L')$ by
tracing it from left to right.  Finding the leftmost edge (ray)
of the level is easy to do in linear time.  Suppose that we are
currently at some point $q$ on some edge $e$ of the level, and let $i$
be the index for which the line $\ell$ containing $e$ belongs to
$L'_i$. The right endpoint $q'$ of $e$ is the nearest intersection of
the rightward-directed ray emanating from $q$ along $e$ with another
line of $L'$.  
We find $q'$ using the dynamic half-space intersection data structure of Overmars and van Leeuwen~\cite{ol-81}. This data structure maintains the intersection of half-spaces under insertions and deletions and supports ray-shooting queries from any point inside the intersection.  The intersection must be non-empty at all times and the ray-shooting query returns the half-space first hit by the ray. We use the data structure as follows: For each $j\neq i$, the face of $\A(L'_j)$ that
contains $q$ contributes the at most four half-spaces defining the face. For $L'_i$, $e$ bounds two faces of
$\A(L'_i)$, the union of which is defined by at most four half-spaces in $L'_i$. We maintain the collection of the at most $4k$ such half-spaces. Each ray-shooting query takes $O(\log^2 k)$ time and half-spaces can be added and removed in the same time bound.

After we obtain $q'$, the new edge $e'$ that the level follows lies on
the new line $\ell'$ containing $q'$ (note that $\ell'$ is unique since our lines are assumed to be in general position); let $j$ be the index for which
$\ell'\in L'_j$. Consider the case $i\neq j$; the case $i=j$ is easier
to handle. For every index $m\ne i,j$, both $q$ and $q'$ lie in the
same face of $\A(L'_m)$, so the at most four lines of $L'_m$ that are
stored in the structure do not change.
For $L'_i$, $e'$ enters one of the two faces of $\A(L'_i)$ adjacent to
$e$. We insert $\ell$ into the structure and delete the opposite line
bounding the other face. For $L'_j$, we are now tracing (along $e'$)
the common boundary of two faces. We delete $\ell'$ from the structure
and insert the line bounding the opposite edge of the new face. 
That is, each new vertex on the $k$\th level takes $O(\log^2k)$ time
to obtain. Since the complexity of the $k$\th (upper) level in an
arrangement of $n$ lines (in general position) is $O(nk^{1/3})$~\cite{d-98}, the total cost
of constructing the level is $O(nk^{1/3}\log^2k)$.

%
%

In conclusion, one can compute the $k$\th upper level $\LUp_k$ of
$\A(L)$ in $O(n \log n + n k + nk^{1/3}\log^2k ) = O(n \log n + nk)$
time.

We come to the final step. We construct the lower convex hull $C_k$ of
$\LUp_k$, which can be done in linear time, that is, in $O(nk^{1/3})$
time, since the vertices of $\LUp_k$ are already sorted from left to
right. Note that each point $q$ on or above $C_k$ lies at upper level
at most $2k$, because every line that passes above $q$ must pass above
at least one of the two endpoints of the edge of $C_k$ that contains
$q$ or passes below $q$. For each line $\ell\in L$ we compute its (one
or two) intersection points with $C_k$, in $O(\log n)$ time, and
thereby obtain its portion above $C_k$.  The overall time for this
step is $O(n\log n + nk^{1/3})$.

Let $S$ denote the resulting collection of at most $n$ segments and
rays.  Since all the elements of $S$ are contained in the at-most-$2k$
upper level of $\A(L)$, the complexity of $\A(S)$ is $O(nk)$
(see~\cite{ag-86}).  We construct $\A(S)$ using the deterministic
algorithm of Chazelle and Edelsbrunner~\cite{ce-oails-92}, which runs
in $O(n\log n+nk)$ time.\footnote{The algorithm~\cite{ce-oails-92}
   runs in $O(n\log n+I)$ time, where $n$ is the number of segments
   and $I$ is the number of intersections that they induce. The same
   holds, in expectation, for the randomized algorithm that we
   cite~\cite{bcko-08}.}  Alternatively, we can use the randomized
incremental algorithm described in \cite{bcko-08}, which runs in expected time
$O(n\log n + nk)$.  Finally, we sweep $\A(S)$ once more to remove any
vertex or edge of the arrangement that lies below $\LUp_k$.  This step
can also be performed in $O(n\log n+nk)$ time, by traversing the
planar map obtained from the previous construction, updating the level
in $O(1)$ time when we cross from one feature to an adjacent one.
$\Box$


\begin{thebibliography}{dBCKO08}

\bibitem[AG86]{ag-86}
\hrefb{http://www.math.tau.ac.il/~nogaa/}{N.~{Alon}} and E.~Gy{\H o}ri.
\newblock The number of small semispaces of a finite set of points in the
  plane.
\newblock {\em J. Combin. Theory Ser. A}, 41:154--157, 1986.

\bibitem[CE92]{ce-oails-92}
\hrefb{http://www.cs.princeton.edu/~chazelle/}{B.~{Chazelle}} and \hrefb{http://www.cs.duke.edu/~edels/}{H.~{Edelsbrunner}}.
\newblock An optimal algorithm for intersecting line segments in the plane.
\newblock {\em \hrefb{http://www.acm.org/jacm/}{J. Assoc. Comput. {Mach.}}}, 39:1--54, 1992.

\bibitem[CS89]{cs-89}
\hrefb{http://cm.bell-labs.com/who/clarkson/}{K.~L. {Clarkson}} and P.~W. Shor.
\newblock Applications of random sampling in computational geometry, {II}.
\newblock {\em \hrefb{http://link.springer.com/journal/454}{Discrete Comput. {}Geom.}}, 4(5):387--421, 1989.

\bibitem[dBCKO08]{bcko-08}
\hrefb{http://www.win.tue.nl/~mdberg/}{M.~de~{Berg}}, \hrefb{http://www.win.tue.nl/~ocheong}{O.~{Cheong}}, {M. van} Kreveld, and \hrefb{http://www.cs.uu.nl/people/markov/}{M.~H. {Overmars}}.
\newblock {\em Computational Geometry: Algorithms and Applications}.
\newblock Springer-Verlag, Santa Clara, CA, USA, 3rd edition, 2008.

\bibitem[Dey98]{d-98}
T.~K. Dey.
\newblock Improved bounds for planar {$k$}-sets and related problems.
\newblock {\em \hrefb{http://link.springer.com/journal/454}{Discrete Comput. {}Geom.}}, 19(3):373--382, 1998.

\bibitem[ERK96]{erk-96}
H.~Everett, J.-M. Robert, and {M. van} Kreveld.
\newblock An optimal algorithm for the {$({\leq} k)$}-levels, with applications
  to separation and transversal problems.
\newblock {\em Int. J. Comput. Geom. Appl.}, 6(3):247--261, 1996.

\bibitem[EW86]{DBLP:journals/siamcomp/EdelsbrunnerW86}
\hrefb{http://www.cs.duke.edu/~edels/}{H.~{Edelsbrunner}} and E.~Welzl.
\newblock Constructing belts in two-dimensional arrangements with applications.
\newblock {\em {SIAM} J. Comput.}, 15(1):271--284, 1986.

\bibitem[HS18]{hs-18}
D.~Halperin and \hrefb{http://www.math.tau.ac.il/~michas}{M.~{Sharir}}.
\newblock Arrangements.
\newblock In J.~E. Goodman, \hrefb{http://cs.smith.edu/~orourke/}{J.~{O'Rourke}}, and Cs.~D. T\'oth, editors, {\em
  Handbook of Discrete and Computational Geometry}, chapter~28, pages 723--762.
  Chapman \& Hall/CRC, Boca Raton, FL, 3rd edition, 2018.

\bibitem[OvL81]{ol-81}
\hrefb{http://www.cs.uu.nl/people/markov/}{M.~H. {Overmars}} and J.~van Leeuwen.
\newblock Maintenance of configurations in the plane.
\newblock {\em J. Comput. Syst. Sci.}, 23:166--204, 1981.

\bibitem[SA95]{sa-95}
\hrefb{http://www.math.tau.ac.il/~michas}{M.~{Sharir}} and \hrefb{http://www.cs.duke.edu/~pankaj}{P.~K.~{Agarwal}}.
\newblock {\em {Davenport-Schinzel} Sequences and Their Geometric
  Applications}.
\newblock Cambridge University Press, New York, 1995.

\bibitem[TT03]{tt-03}
H.~Tamaki and T.~Tokuyama.
\newblock A characterization of planar graphs by pseudo-line arrangements.
\newblock {\em Algorithmica}, 35(3):269--285, 2003.

\end{thebibliography}
\end{document}